\documentclass[
reprint,
superscriptaddress,
amsmath,
amssymb,  
]{revtex4-2}

\usepackage{graphicx}
\usepackage{dcolumn}
\usepackage{bm}
\usepackage{hhline}
\usepackage[switch]{lineno}
\usepackage{array}
\usepackage{xcolor}
\usepackage{soul}
\usepackage[indent,skip=6pt]{parskip}

\begin{document}

\title{Electronic and Photonic Integration of Single Quantum Emitters in 2D Materials}



\author{Sahil D. Patel}
\altaffiliation{These authors contributed equally to this work.}
\affiliation{\mbox{Electrical and Computer Engineering Department, University of California, Santa Barbara, CA 93106, USA}}

\author{Sean Doan}
\altaffiliation{These authors contributed equally to this work.}
\affiliation{Physics Department, University of California, Santa Barbara, CA 93106, USA}

\author{Luka Jevremovic}
\affiliation{\mbox{Electrical and Computer Engineering Department, University of California, Santa Barbara, CA 93106, USA}}

\author{Kamyar Parto}
\affiliation{\mbox{Electrical and Computer Engineering Department, University of California, Santa Barbara, CA 93106, USA}}

\author{Galan Moody}
\email{moody@ucsb.edu}
\affiliation{\mbox{Electrical and Computer Engineering Department, University of California, Santa Barbara, CA 93106, USA}}

\date{\today}


\begin{abstract}

Single-photon sources that are simultaneously bright, pure, and interference-ready are a cornerstone technology for quantum communication and photonic quantum information processing. Yet many state-of-the-art solid-state emitters still rely on laboratory-style operation requiring bulky optical excitation, careful alignment, and extensive post-selection to find operating points with acceptable linewidth, stability, and brightness. Scalable quantum photonics instead demands turnkey quantum-light engines that can be triggered on-demand, stabilized against environmental noise, and efficiently interfaced with fibers and photonic circuitry, especially when moving from a single optimized device to arrays where device variability, packaging overhead, and long-term drift dominate system complexity. This review covers recent progress in electronic and photonic integration of single quantum emitters in two-dimensional (2D) materials, organized around the premise that these approaches address complementary bottlenecks on the path to deployable sources. On the electronic side, we survey device concepts that enable triggered operation through electrical injection and fast modulation, as well as electrostatic stabilization and tunability that suppress charge-noise-induced spectral wandering and blinking, narrowing emission lines and improving reproducibility. On the photonic side, we review how integrated waveguides and resonators reshape the electromagnetic environment to funnel emission into a single collectable mode and, when desired, increase radiative rates via Purcell enhancement, thereby boosting usable photon flux and relaxing downstream optical-loss constraints. Focusing on localized excitonic emitters in transition metal dichalcogenides and defect-based color centers in hexagonal boron nitride, we connect materials physics to the device-level metrics that are most pertinent for quantum technologies such as single-photon purity, brightness into a well-defined mode, spectral stability, and ultimately photon indistinguishability. We position that the next phase of progress will be driven by intentional co-design of electronics and photonics, and we outline integrated architectures that jointly optimize on-demand operation, stabilization and tunability, and packaging-compatible optical interfacing.

\end{abstract}

 

\maketitle
\section*{Introduction}
Single-photon sources that are simultaneously bright, pure, and interference-ready are a central enabling technology for quantum communication and photonic quantum information processing. Yet many of the most compelling demonstrations of solid-state quantum emitters remain at the laboratory setting such that they rely on table-top optical excitation, careful optical alignment, and extensive post-selection to find operating points where linewidth, stability, and collection efficiency are acceptable. In practice, scalable quantum photonics will require turnkey quantum-light engines that can be triggered on demand, stabilized against environmental noise, and efficiently interfaced with fibers and photonic circuitry. This need becomes even more crucial when moving from a single optimized device to arrays of sources, where device-to-device variability, packaging overhead, and long-term drift can dominate system complexity.

\begin{figure*}[t!] \centering
     \includegraphics[scale=0.65]{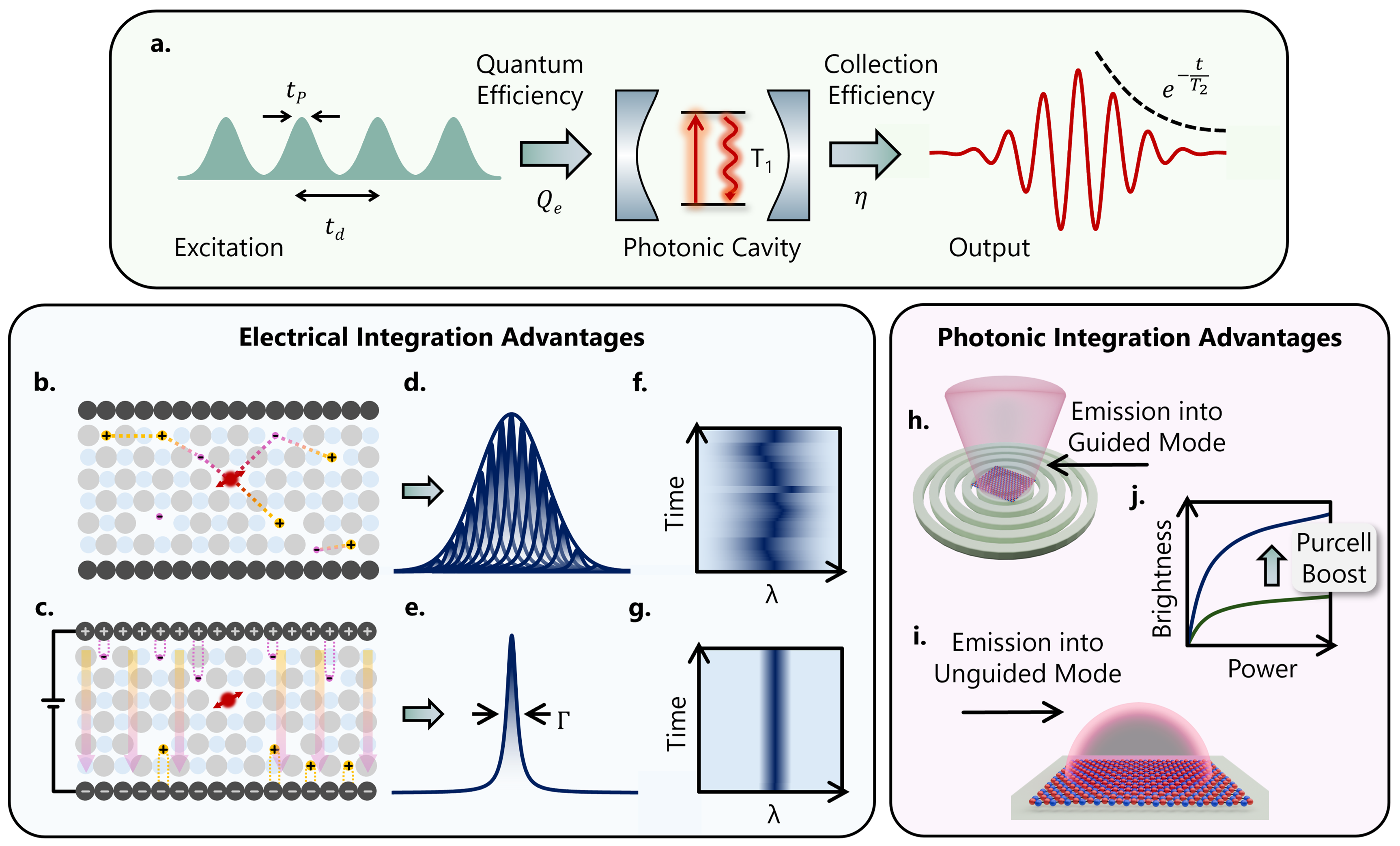}  
     \vspace{-0.5cm}
          \caption{\footnotesize \textbf{Electronic stabilization and photonic enhancement of 2D quantum emitters.} \textbf{(a)} An input excitation train---delivered either optically or electrically---consists of pulses with temporal width $t_p$ and repetition period $t_d$. Excitation is converted into an emissive state with quantum efficiency $Q_e$ and coupled to an on-chip photonic cavity, where the radiative lifetime $T_1$ sets the characteristic emission timescale. The generated photons are extracted and routed with an overall collection efficiency $\eta$, yielding an output field whose temporal coherence is captured by the coherence time $T_2$ (envelope $\propto e^{-t/T_2}$). \textbf{(b)} A defect-based quantum emitter (red) embedded in a solid-state host experiences nearby fluctuating charge traps and local electric fields, which couple to the defect and introduce dephasing and inhomogeneous broadening. \textbf{(c)} In a gated device architecture, an applied electric field sweeps mobile charges toward electrodes and stabilizes the local electrostatic environment, suppressing charge-noise-induced decoherence. \textbf{(d)} In the presence of charge noise, the emission line is broadened due to stochastic Stark shifts and additional dephasing channels. \textbf{(e)} Electrostatic gating reduces the dynamic charge fluctuations, yielding a narrower linewidth and improved optical coherence. \textbf{(f)} Fluctuating charges lead to spectral hopping and wandering, producing fast and slow timescale spectral diffusion (temporal jitter and drift of the emission wavelength). \textbf{(g)} With electrostatic stabilization in a gated device, the emission wavelength remains stable in time, enabling spectrally robust single-photon generation. \textbf{(h)} Coupling a 2D quantum emitter to a photonic cavity or guided-mode structure funnels emission into a well-defined optical mode, enabling highly efficient (potentially near-unity) collection and routing of the emitted photons. \textbf{(i)} In the absence of photonic engineering, emission radiates into uncontrolled, unguided modes within a locally inhomogeneous dielectric environment, limiting the experimentally collected fraction due to finite numerical aperture and the inability to capture light emitted into all directions. \textbf{(j)} Resonant photonic confinement can also enhance the radiative spontaneous emission rate via the Purcell effect, increasing detected count rates and brightness at a given excitation power.  Together, improved mode engineering (higher collection efficiency) and Purcell-enhanced emission (higher radiative rate) increase source efficiency and enable brighter, higher-rate single-photon generation for scalable quantum photonics.
}
          \label{fig: intro}
\end{figure*}

The central theme of this review is that electronic integration and photonic integration address complementary bottlenecks on the path to scalable quantum-light sources. Figure~\ref{fig: intro} summarizes this perspective. At the most basic level (Fig.~\ref{fig: intro}a), an excitation train, delivered optically or electrically, prepares an emissive state with quantum efficiency $Q_e$, which then relaxes with a characteristic radiative lifetime $T_1$. To be useful in scalable architectures, the emitted photons must be efficiently extracted and routed (overall efficiency $\eta$), while maintaining coherence quantified by $T_2$. In solid-state environments, the emitter is embedded in a fluctuating electrostatic landscape that can degrade $T_2$, broaden the emission line, and introduce spectral wandering (Fig.~\ref{fig: intro}). Electronic integration provides a direct route to control this environment, while photonic integration provides a route to engineer the optical mode into which the photon is emitted. Motivated by this division of roles, we first outline the device-level requirements for scalable quantum-light sources and then introduce the two primary 2D material emitter platforms considered throughout this review.

\subsection{Scalable and Turnkey Quantum Light Sources}

A practical single-photon source must satisfy a set of coupled requirements: (i) triggered operation with a known emission time relative to an input clock, (ii) high single-photon purity (low multiphoton probability), (iii) high brightness into a single optical mode to enable low-loss collection and routing, and (iv) spectral stability and optical coherence to enable interference between photons produced at different times and from different single quantum emitter (SQE). These figures of merit are not independent. For example, increasing repetition rate demands short effective emission timescales (often set by $T_1$), while achieving high indistinguishability additionally requires suppressing dephasing mechanisms that reduce $T_2$. Likewise, even an optically coherent SQE is not system-ready if most emission radiates into unguided modes and cannot be efficiently extracted.

\vspace{-5pt}
\subsubsection{Optical Excitation vs Electrical Injection}

Most demonstrations of 2D-material quantum emitters begin with optical excitation \cite{fournier2021position, ahmed2025nanoindentation, doan2025near,he2015single, srivastava2015optically, koperski2015single, tonndorf2015single}, where a pulsed or continuous-wave laser prepares an excitonic or defect-bound emissive state. In the pulsed case (Fig.~\ref{fig: intro}a), the excitation sequence is characterized by a pulse width $t_p$ and repetition period $t_d$, which set the temporal structure of photon generation. Optical excitation is experimentally convenient and can access resonant or quasi-resonant pumping schemes that reduce background fluorescence and improve coherence, but it also introduces substantial system overhead such as stable laser delivery, optical alignment, filtering, and for resonant excitation, suppression of scattered pump light.

Electrical injection provides an alternative pathway to triggered operation that is naturally compatible with chip-scale architectures. In an electrically driven source, the excitation is delivered through an integrated junction or tunneling structure that injects carriers into the emissive region, ideally creating the same localized radiative state without requiring an external laser. In the turnkey limit, electrical drive reduces the optical footprint of the system and enables direct electronic modulation, addressing a key scalability constraint when moving from one SQE to arrays. At the same time, electrical injection must be engineered to preserve quantum emission where the injected carrier environment can introduce charge noise, excess heating, and background electroluminescence that degrade purity and coherence unless the device architecture tightly localizes recombination and stabilizes the SQE's charge configuration.

\vspace{0pt}
\subsubsection{Electrical Control and Tunability}

Even under purely optical excitation, the local electrostatic environment often sets the dominant limitation to coherence and reproducibility. A defect or localized exciton embedded in a solid-state host experiences nearby fluctuating charges and local electric fields (Fig.~\ref{fig: intro}b), which can imprint stochastic Stark shifts and additional dephasing channels that broaden the emission spectrum (Fig.~\ref{fig: intro}d) and cause temporal spectral diffusion (Fig.~\ref{fig: intro}f). These processes degrade photon indistinguishability and can prevent reliable spectral alignment between nominally similar SQEs.

Electrostatic gating converts this uncontrolled environment into a controllable device parameter. By sweeping mobile charges toward electrodes and controlling trap occupation (Fig.~\ref{fig: intro}c), gating can suppress charge-noise-induced decoherence, narrowing the optical linewidth (Fig.~\ref{fig: intro}e) and stabilizing the emission frequency in time (Fig.~\ref{fig: intro}g). Importantly, the same electrodes that stabilize the environment can also tune the transition energy via the quantum-confined Stark effect, enabling spectral matching between SQEs and/or alignment to narrowband photonic resonances. In this sense, ``electrical control'' should be viewed not as an optional add-on, but as a key enabling route to transforming a stochastic, drifting SQE into a reproducible component suitable for interference-based quantum photonics.

\vspace{-3pt}
\subsubsection{Photonic Integration}

While electrical control targets coherence and stability, photonic integration targets mode engineering. In a homogeneous dielectric environment without photonic structures, an SQE radiates into a continuum of unguided modes (Fig.~\ref{fig: intro}i), and only a small fraction is captured by a finite numerical aperture or by incidental coupling pathways. Integrated photonics instead reshapes the electromagnetic environment so that emission is preferentially funneled into a desired channel. Coupling to a guided mode or an engineered outcoupler (Fig.~\ref{fig: intro}h) enables efficient collection and routing on chip, and provides a natural interface to filtering, switching, and multiplexing elements that are essential for scale-up.

Resonant photonic confinement additionally enables radiative-rate control via the Purcell effect (Fig.~\ref{fig: intro}j), which can increase brightness at a given excitation level and shorten the effective emission timescale. Together, enhanced extraction (higher $\eta$) and Purcell-enhanced emission (higher radiative rate) increase the usable photon flux and reduce the effective cost of every downstream optical loss mechanism. These advantages motivate the organization of the photonic-integration sections in this review into three experimentally distinct but conceptually related strategies: waveguide integration for broadband routing, on-chip cavity integration for enhancement and circuit compatibility, and off-chip cavity integration for high directionality into free space and fiber-compatible modes.

\subsection{2D Material Quantum Emitters}

\begin{figure*}[t!] \centering
     \includegraphics[scale=0.9]{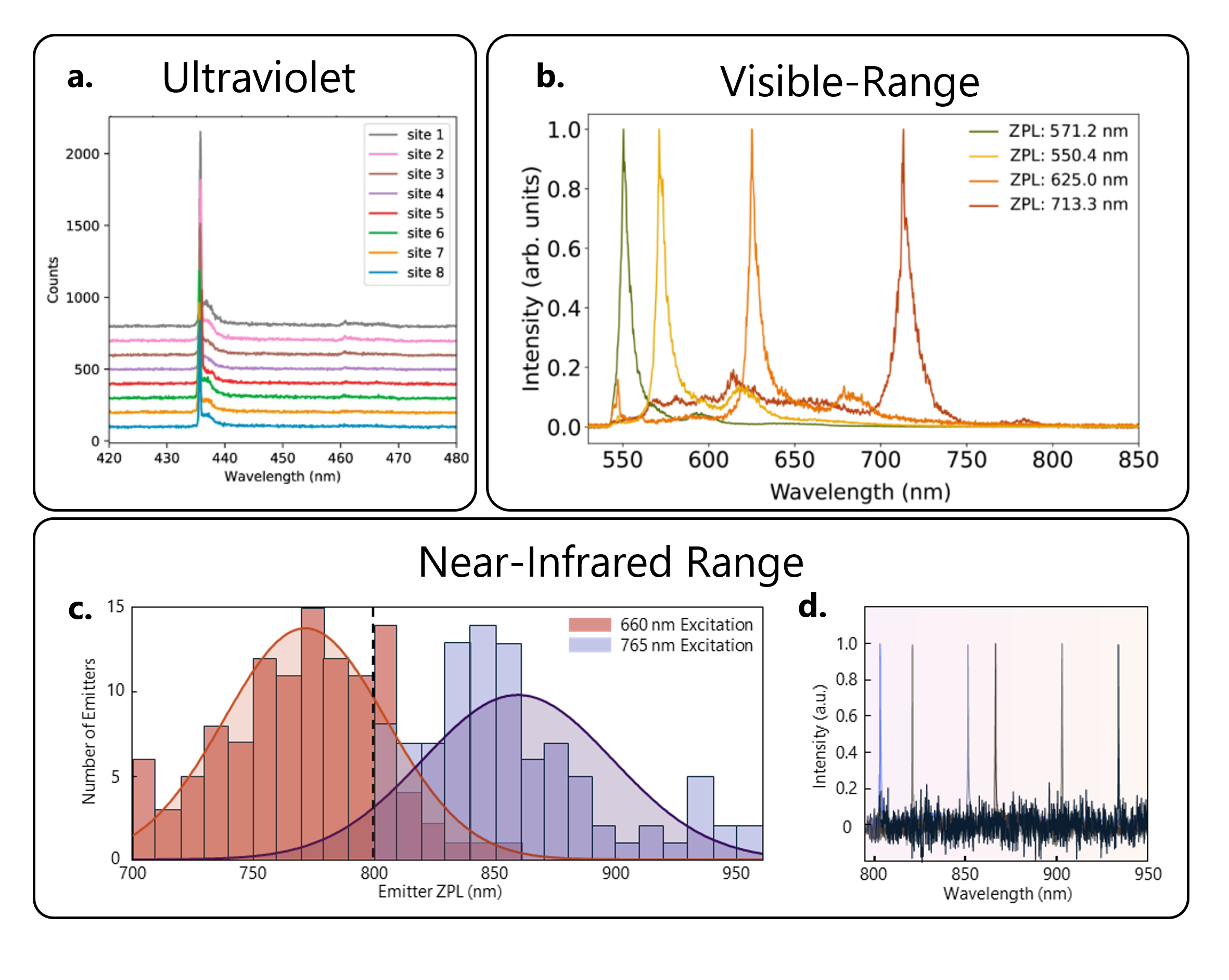}  
     \vspace{-0.8cm}
          \caption{\footnotesize\textbf{Spectral diversity of quantum emitters in hBN across ultraviolet, visible, and near-infrared wavelength bands.} \textbf{(a)} Representative ultraviolet photoluminescence spectra from multiple sites, illustrating site-to-site variability in UV emission features (adapted from \cite{fournier2021position} under the CC BY 4.0 license). \textbf{(b)} Visible-range emission spectra from distinct emitters, plotted and labeled by their zero-phonon line (ZPL) wavelength, highlighting a broad distribution of discrete ZPL positions across the visible band (adapted from \cite{ahmed2025nanoindentation}, Copyright 2025, American Chemical Society). \textbf{(c)} Histogram of NIR emitter ZPL wavelengths measured under two excitation conditions (660 nm and 765 nm) at cryogenic temperatures, showing excitation-dependent sampling of the emitter population and a widened spread of operating wavelengths in the NIR (adapted from \cite{doan2025near} under the CC BY 4.0 license). \textbf{(d)} Broadband cryogenic spectrum showing multiple narrow NIR ZPL lines across the 800--950 nm range, illustrating both the discrete spectral landscape and the accessibility of multiple channels for wavelength-selective targeting (adapted from \cite{doan2025near} under the CC BY 4.0 license). }
          \label{fig: hbn emitters}
\end{figure*}

\subsubsection{Transition Metal Dichalcogenides}

Semiconducting transition metal dichalcogenide (TMD) monolayers, most prominently tungsten diselenide (WSe$_2$), provide a route to quantum emission in which the relevant optical transitions are excitonic and inherit the strong light-matter interaction, spin-valley structure, and gate-tunability of atomically thin direct-gap semiconductors. In the delocalized regime, excitons in WSe$_2$ already exhibit valley-selective optical selection rules and pronounced many-body physics, and these same ingredients carry over when carriers are confined to nanoscale potentials. In particular, sharp and spectrally isolated emission lines can emerge when excitons become trapped in confinement landscapes created by local strain gradients, disorder, or point-like defects, producing ``0D-like'' localized excitons that exhibit photon antibunching and other SQE signatures at cryogenic temperatures \cite{he2015single, srivastava2015optically, koperski2015single, tonndorf2015single}. These localized SQEs therefore connect two useful perspectives as they can be viewed as quantum-dot-like excitonic states formed within an atomically thin semiconductor, while still retaining the device-friendly attributes of a vdW material platform, including straightforward stacking into heterostructures and compatibility with electrostatic gating.

A key practical feature of WSe$_2$ is that the confinement landscape is highly engineerable, enabling a tangible route from random spatially generated SQEs to designed SQE sites. Localized lines can occur naturally at flake edges and interfaces \cite{koperski2015single}, but can also be created reproducibly by imposing controlled strain fields using nanopillars, nanoindentation, or related topographic templates, which enables deterministic positioning and scalable arrays of SQEs \cite{branny2017deterministic, palaciosberraquero2017large, kern2016nanoscale, parto2021defect}. Beyond templated strain, ``self-assembled'' strain features such as nanobubbles and wrinkles generate localized exciton states in WSe$_2$ and related heterostructures, helping clarify how strain and disorder cooperate to form quantum-emitter potentials \cite{shepard2017nanobubble, darlington2020imaging}. Together, these demonstrations make TMD emitters a natural entry point for 2D quantum photonics. They combine exciton localization in engineered potentials, analogous to semiconductor quantum dots for creating quantum emission, but with intrinsic compatibility with layered heterostructures and electrostatic control, providing a direct pathway toward site-controlled single-photon emission through strain-engineered localization.

\subsubsection{Hexagonal Boron Nitride}

\begin{figure*}[t!] \centering
     \includegraphics[scale=0.55]{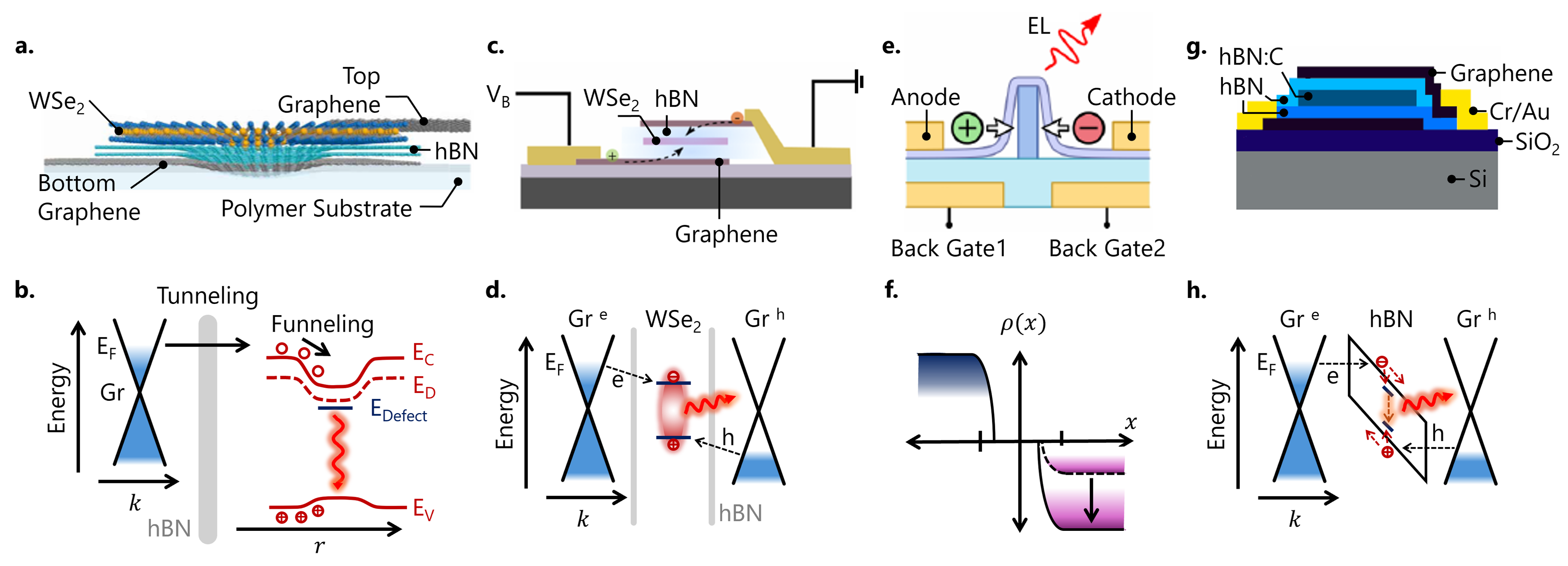}  
     \vspace{-0.5cm}
          \caption{\footnotesize\textbf{Electrical Injection Theory and Device Concepts.} \textbf{(a)} Schematic of a representative vertical vdW heterostructure used for electrical injection into TMD emitters. A monolayer WSe$_2$ contacted by top and bottom graphene electrodes, with a thin hBN spacer serving as a tunneling barrier to enable bias-controlled carrier injection and electroluminescence from the TMD layer (adapted from \cite{so2021electrically} under the CC BY 4.0 license). \textbf{(b)} Illustrative band-diagram of the graphene/hBN/TMD stack under bias, showing carrier tunneling from graphene through the hBN barrier into the TMD bands, followed by energy funneling into a localized emissive state that supports radiative recombination and electrically driven quantum emission. \textbf{(c)} Schematic of a second vdW heterostructure device concept for electrically driven emission in a monolayer TMD. WSe$_2$ is sandwiched between thin hBN tunnel barriers and contacted by metal electrodes to the graphene layers, enabling bias-controlled vertical carrier injection into the TMD (adapted from \cite{clark2016single}, Copyright 2016, American Chemical Society). \textbf{(d)} Illustrative band diagram under applied bias showing electron and hole tunneling from the two graphene electrodes through the hBN barriers into WSe$_2$, followed by carrier capture/exciton formation (including into localized defect-bound states) and electroluminescent emission from radiative recombination in the TMD. \textbf{(e)} Schematic of a monolayer TMD device in which two independently biased back gates electrostatically define adjacent p- and n-doped regions, forming a narrow lateral p--n junction between anode/cathode contacts where electrons and holes are injected from opposite sides and electroluminescence is spatially confined to the junction and can be aligned with a site-engineered localized SQE (adapted from \cite{lenferink2022tunable}, Copyright 2022, American Chemical Society). \textbf{(f)} Illustrative profile of the lateral carrier/charge density $\rho$(x) and corresponding potential/band bending across the split-gate junction, showing a steep density gradient and a junction region that promotes electron--hole overlap, exciton formation, and radiative recombination, thereby localizing EL to a designed position. \textbf{(g)} Representative vertical van der Waals heterostructure used to electrically drive hBN quantum emitters. A hBN layer hosting color centers/quantum emitters embedded within an hBN tunnel barrier and sandwiched between graphene electrodes, with metal contacts to graphene on a Si/SiO$_2$ substrate (adapted from \cite{grzeszczyk2024electroluminescence} under the CC BY 4.0 license). \textbf{(h)} Illustrative band diagram showing how an applied vertical bias shifts the graphene chemical potentials to enable electron/hole tunneling into localized defect levels in hBN layer hosting defects, followed by radiative recombination via intradefect or defect--band transitions, producing electrically driven emission from hBN color centers.}
          \label{fig: electrical theory}
\end{figure*}

Color centers in hBN are also a compelling solid-state platform for quantum light generation and, increasingly, spin--photon functionality, which are capabilities that underpin quantum communication links, networked quantum nodes, and photonic quantum information processing \cite{aharonovich2016solid, o2009photonic, aharonovich2022quantum}. As a wide-bandgap vdW semiconductor, hBN combines chemical and thermal robustness with device practicality, enabling stable operation under ambient conditions in a manner often associated with more established bulk hosts such as diamond \cite{kianinia2017robust}. Its large bandgap ($\sim$6 eV) supports a rich set of optically active defect configurations, with emission spanning from the ultraviolet (UV) through the visible and into the near-infrared (NIR) \cite{cassabois2016hexagonal, caldwell2019photonics, fournier2021position, gale2022site, gottscholl2020initialization, doan2025near}. This wavelength-scale diversity is summarized in Fig.~\ref{fig: hbn emitters}, spanning UV B-centers (Fig.~\ref{fig: hbn emitters}a) \cite{fournier2021position}, visible carbon-related emitters (Fig.~\ref{fig: hbn emitters}b) \cite{ahmed2025nanoindentation}, and emerging NIR defects (Fig.~\ref{fig: hbn emitters}c, d) \cite{doan2025near}. In these systems, spatially localized electronic states introduce optical transitions within the bandgap, enabling efficient generation of quantum light from atom-like emitters embedded in a crystalline host \cite{tran2016quantum, tran2016robust, jungwirth2016temperature}.

Beyond the intrinsic materials advantages, the layered geometry of hBN introduces a unique 2D host for color centers. Defects can reside within an atomically thin membrane while remaining optically active near surfaces and interfaces, supporting strong proximity coupling to the surrounding environment and facilitating electrostatic, strain, and photonic engineering in compact device stacks \cite{grosso2017tunable, luo2018deterministic, zhao2021site}. This same geometry can also simplify photon extraction and collection compared to deeply embedded bulk SQEs \cite{tran2016quantum, tran2016robust, jungwirth2016temperature, grosso2017tunable}. Equally important, hBN has demonstrated scalable routes toward site-controlled and deterministic SQE formation, which is an essential step toward manufacturable quantum-light-source arrays \cite{fournier2021position, gale2022site, gan2022large, li2021scalable, xu2021creating}. Combined with straightforward transfer and heterogeneous integration, these attributes make hBN naturally compatible with mature electronic and photonic platforms \cite{o2024transfer, froch2020coupling, li2021integration, parto2022cavity, patel2024surface, sakib2024purcell, yamashita2025deterministic, kim2018photonic}. A growing set of hBN defects exhibit clear spin--optical signatures from cryogenic to room temperature, motivating hBN not only as a bright single-photon source material, but also as a realistic candidate for spin--photon interfaces and quantum sensing modalities \cite{gottscholl2020initialization, stern2022room, patel2024room, chejanovsky2021single, gao2022nuclear, gao2021high}.

At the level of specific defect families, carbon-related single quantum defects in hBN (emitting in the visible) have been shown to host optically addressable spin states that can be accessed through optically detected magnetic resonance (ODMR) (as illustrated by the representative visible-range zero-phonon line (ZPL) distribution in Fig.~\ref{fig: hbn emitters}b)
 \cite{stern2022room, patel2024room, chejanovsky2021single, stern2024quantum}. In the UV, the so-called B-center represents a distinct and highly promising class where these SQEs exhibit strong spectral homogeneity (around 436 nm; see Fig.~\ref{fig: hbn emitters}a) and narrow cryogenic linewidths, enabling key multi-emitter benchmarks such as the first resolvable two-photon interference dip reported for hBN \cite{fournier2023two}. Notably, B-centers also support comparatively high-yield, deterministic generation which is an attribute that is difficult to achieve across many other hBN defect complexes \cite{fournier2021position, gale2022site}. Native boron-vacancy-related defects, particularly $V_B^-$, have also attracted attention as spin-active centers with NIR ensemble emission and measurable ODMR contrast \cite{gottscholl2020initialization, gottscholl2021room, baber2021excited, alzahrani2024negatively, robertson2023detection, gong2023coherent}. However, despite their promise for spin physics and sensing, reliably isolating single $V_B^-$ emitters remains a central challenge, limiting near-term applicability for deterministic single-photon-source implementations. Complementing these established visible and UV families, recently identified oxygen-related defects extend hBN quantum emission into the NIR as reflected by the ZPL statistics and example spectra in Fig.~\ref{fig: hbn emitters}c, d \cite{doan2025near} with ultra-high single photon purity up to 99.9$\%$ with $>1$ MHz rates.

\section*{Electrical Addressability}

Electrical addressability provides a scalable route to operate 2D material quantum emitters as turnkey quantum-light sources, where photons are generated and stabilized using device-level control. By integrating contacts, gates, and dielectric stacks, electrical platforms enable on-demand excitation through current injection, fast intensity modulation, and programmable selection of emitting sites, while simultaneously providing a direct handle on the local electrostatic environment that governs charge noise, blinking, and spectral diffusion. Beyond simply turning emission on, electrostatic control enables Stark tuning, charge-state preparation, and suppression of non-radiative pathways---capabilities that become especially critical when moving from single optimized emitters to large arrays that require reproducibility, spectral alignment, and long-term stability. The following sections survey this electrical toolbox beginning with electrically injected emission in TMDs and hBN, and then turning to electrical tunability for Stark tuning and charge control.

\subsection{Electrical Injection}

\begin{figure*}[t!] \centering
     \includegraphics[scale=0.85]{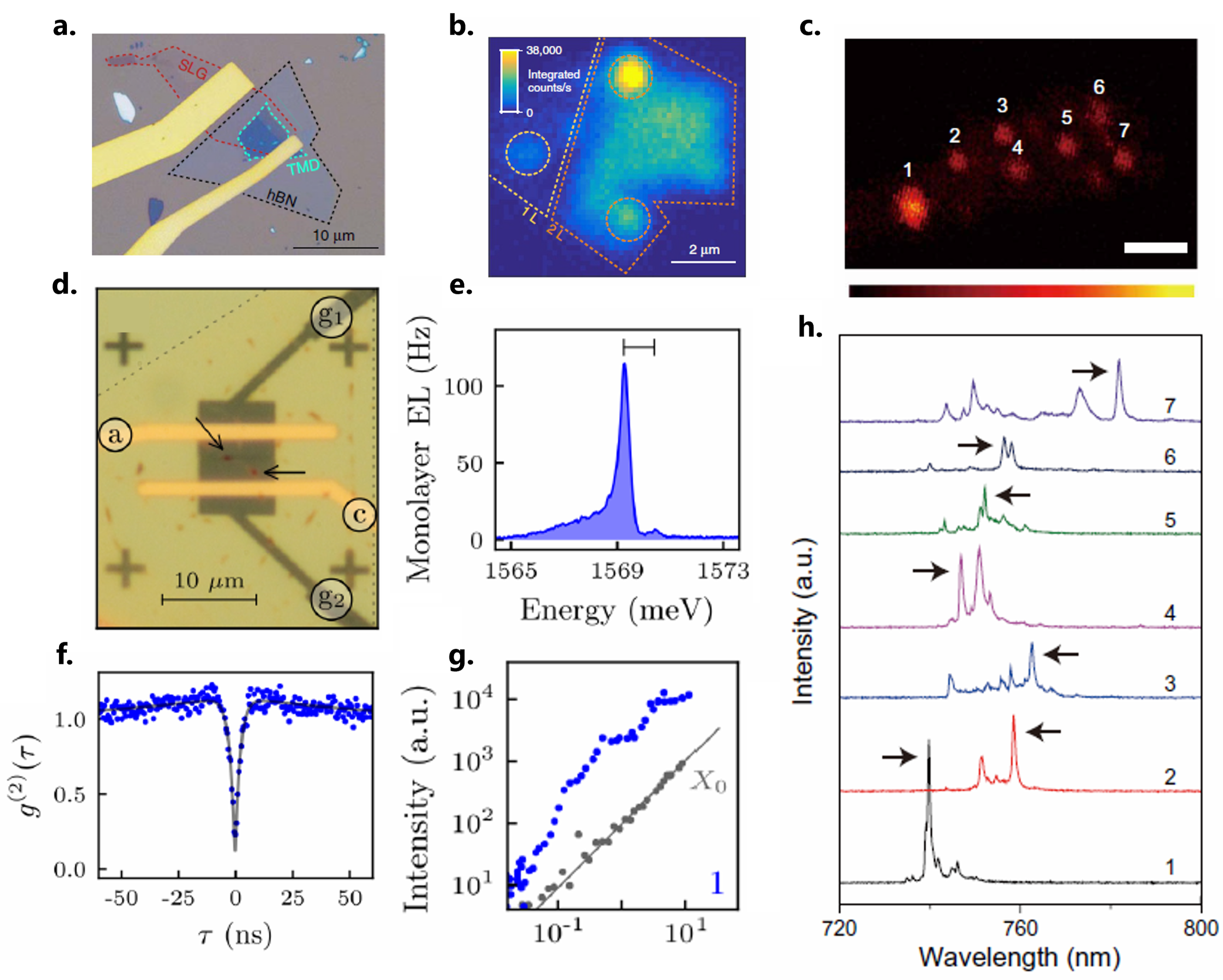}  
     \vspace{-0.5cm}
          \caption{\footnotesize \textbf{Electrical injection of quantum emitters in atomically-thin TMD--based QLEDs.} \textbf{(a)} Vertical graphene/hBN/WSe$_2$ tunnel-junction LED device used to electrically drive emission in monolayer WSe$_2$ (adapted from \cite{palacios2018atomically} under the CC BY 4.0 license). \textbf{(b)} Representative spatially localized electroluminescence (EL) intensity map from the same device class, highlighting sub-micron EL hotspots under electrical drive (adapted from \cite{palacios2018atomically} under the CC BY 4.0 license). \textbf{(c)} Localized EL imaging in a vertical vdW injection platform illustrating multiple electrically addressable emitting sites within the active region (adapted from \cite{so2021electrically} under the CC BY 4.0 license). \textbf{(d)} Split-gate lateral p--n junction device concept for electrically driven localized excitons, where independent gates define adjacent p- and n-type regions and confine recombination to a narrow junction (adapted from \cite{lenferink2022tunable}, Copyright 2022, American Chemical Society). \textbf{(e)} Representative EL spectrum dominated by a localized excitonic feature in the gate-defined junction geometry (adapted from \cite{lenferink2022tunable}, Copyright 2022, American Chemical Society). \textbf{(f)} Second-order intensity autocorrelation under electrical pumping demonstrating antibunching from a localized SQE in the lateral junction platform (adapted from \cite{lenferink2022tunable}, Copyright 2022, American Chemical Society). \textbf{(g)} Current-dependent EL response (and associated background/excitonic contributions) illustrating electrically driven operation and saturation-like behavior of localized emission in the junction geometry (adapted from \cite{lenferink2022tunable}, Copyright 2022, American Chemical Society). \textbf{(h)} Site-resolved EL spectra from a vertical junction platform showing distinct narrow emission lines across different localized emitters shown in \textbf{(c)} (adapted from \cite{so2021electrically} under the CC BY 4.0 license). }
          \label{fig: TMD EL}
\end{figure*}

Electrical injection of 2D material quantum emitters has converged on a small set of device archetypes that reliably deliver carriers to a localized radiative state while retaining electrostatic control of the recombination environment. Figure \ref{fig: electrical theory} summarizes three representative approaches: (i) vertical van der Waals  (vdW) tunnel junctions for TMD emitters (Fig. \ref{fig: electrical theory}a and \ref{fig: electrical theory}b) \cite{so2021electrically}, (ii) double tunnel-barrier vertical LEDs that enable bipolar injection and efficient defect-bound emission in TMDs (Fig. \ref{fig: electrical theory}c and \ref{fig: electrical theory}d) \cite{clark2016single}, and (iii) lateral, gate-defined p-i-n junctions that confine the recombination zone in-plane with tunable electrostatics (Fig. \ref{fig: electrical theory}e and \ref{fig: electrical theory}f) \cite{lenferink2022tunable, grzeszczyk2024electroluminescence}. An analogous tunnel-junction strategy has also emerged for hBN color centers, where emission is driven through defect-mediated transport in a wide-bandgap host (Fig. \ref{fig: electrical theory}g and \ref{fig: electrical theory}h).

A canonical electrical-injection strategy for TMD quantum emitters employs a vertical vdW tunnel junction, where a monolayer TMD (e.g., WSe$_2$) is contacted by top and bottom graphene electrodes and an atomically thin hBN dielectric spacer on one side sets a well-controlled tunneling barrier (graphene/WSe$_2$/hBN/graphene) (see Fig. \ref{fig: electrical theory}a) \cite{so2021electrically}. The central device-level advantage is area-wide, vertically driven injection where the hBN barrier regulates current flow and enables field-assisted tunneling into the monolayer, avoiding the strong contact-edge localization that can occur in lateral devices. In the corresponding band-diagram picture (Fig. \ref{fig: electrical theory}b), bias shifts the graphene chemical potential relative to the TMD bands to permit tunneling injection, after which carriers relax and are captured into a localized strain potential (often framed in terms of bright/dark exciton states plus a defect-associated level), from which narrowband electroluminescence can be produced.

Another closely related variant sandwiches the monolayer TMD between two hBN tunnel barriers contacted by graphene, forming a double-tunnel-barrier vertical LED that supports robust bipolar injection. In the unbiased configuration, the graphene Fermi level lies within the TMD bandgap and both electron and hole injection are energetically suppressed by the barriers. Under applied bias (Fig. \ref{fig: electrical theory}d), the graphene chemical potentials split such that one side favors electron tunneling into the conduction band and potentially sub-gap states while the other favors hole tunneling into the valence band states \cite{clark2016single}. Once both carrier species are present in the monolayer, strong Coulomb interactions promote exciton formation and radiative recombination, yielding electroluminescence that can transition from intrinsic excitonic emission at higher injection to defect-bound exciton emission near threshold.

A second major modality defines the injection region in-plane using a split-gate p-i-n junction within a single monolayer. As shown schematically (Fig. \ref{fig: electrical theory}e), independently biased back gates electrostatically dope adjacent regions p-type and n-type, while source/drain contacts drive lateral current so that electron and hole injection occurs from opposite sides and radiative recombination is confined to the narrow intrinsic undoped junction region. This built-in spatial confinement is particularly attractive for quantum-emitter operation because the junction functions as an electrically defined recombination zone that can be combined with site engineering (e.g., aligning a localized exciton/defect/strain site to the junction) to achieve deterministic, electrically driven emission. The accompanying electrostatics plot (Fig. \ref{fig: electrical theory}f) summarizes the mechanism via the lateral carrier-density profile $\rho(x)$ where the split-gate bias creates a steep charge-density gradient and a junction region where band bending and quasi-Fermi level alignment favor electron--hole overlap and exciton formation \cite{lenferink2022tunable}. Because junction width and field profile are set by gate geometry and the dielectric stack, this approach naturally localizes EL to a designed position while preserving independent gate control over the local carrier density and electrostatic environment during electrical pumping.

Extending these concepts to wide-bandgap defect hosts in hBN, Fig. \ref{fig: electrical theory}g and \ref{fig: electrical theory}h highlight an emerging hBN color-center tunnel-junction LED device. Here, a quantum defect-containing hBN layer is embedded within an hBN stack and sandwiched between graphene electrodes (see Fig. \ref{fig: electrical theory}g) \cite{grzeszczyk2024electroluminescence}. Because hBN is insulating and not doped in the conventional semiconductor sense, electrical excitation proceeds primarily through defect-state-assisted tunneling rather than band-to-band injection (Fig. \ref{fig: electrical theory}h). Under bias, the graphene chemical potentials shift into energetic alignment with mid-gap defect levels, enabling sequential tunneling and capture of electrons and holes into localized states, followed by radiative decay either via intradefect transitions or defect-band-assisted recombination at higher fields. In this framework, the key design knob is the competition between tunneling escape and radiative recombination at the defect, which can be tuned through barrier thickness, interfaces, and electrostatic bias to favor efficient electrically driven color-center quantum emission. With these device concepts in place, the following sections survey the experimental progress enabled by these architectures, focusing on how electrical injection produces electroluminescence from localized defect states in TMDs and hBN, the operating regimes where defect-bound quantum emission dominates, and the performance metrics that determine suitability for electrically driven single-photon sources.

\subsubsection{Transition Metal Dichalcogenides}

Building on the device archetypes introduced above, electrical excitation of TMD-based quantum emitters has progressed along two complementary directions focusing on (i) vertical vdW tunnel junctions that inject carriers through atomically thin barriers into a monolayer TMD, and (ii) lateral electrostatically defined junctions that confine electron-hole overlap to a designed intrinsic in-plane recombination zone. Across both approaches, the central challenge is the same with achieving reliable carrier injection while preserving (or enhancing) recombination through localized excitonic states that yield narrowband, antibunched EL, rather than broad-area free-exciton emission.

A key milestone toward electrically driven quantum emission in monolayer WSe$_2$ was the demonstration of atomically thin quantum light-emitting diodes (QLEDs) based on vertical graphene/hBN/WSe$_2$ heterostructures, where graphene serves as the transparent electrodes and an atomically thin hBN layer provides the tunnel barrier for carrier injection into the monolayer \cite{palacios2018atomically}. In this architecture (see Fig. \ref{fig: TMD EL}a and \ref{fig: TMD EL}b), an applied out-of-plane bias splits the graphene chemical potentials and opens energetically allowed tunneling pathways into the WSe$_2$ band-edge and sub-gap manifolds where injected carriers subsequently relax into localized exciton states (often associated with disorder/defect/strain potentials) that dominate the EL near threshold. Crucially, Palacios-Berraquero \textit{et al.} verified antibunching under electrical drive with $g^{(2)}(0)=0.29\pm0.08$ at $T\approx 4.2$~K, establishing an early quantitative benchmark for electrically pumped single-photon emission in a monolayer semiconductor \cite{palacios2018atomically}. In representative devices, EL operation was obtained at milliampere-scale injection (e.g., EL maps acquired at $I=3$~mA and $V=12.4$~V) (see Fig. \ref{fig: TMD EL}b), while the spatially localized, spectrally sharp emission features remained prominent in the NIR at low injection \cite{palacios2018atomically}. Closely related vertical ``double-barrier'' devices (graphene/hBN/WSe$_2$/hBN/graphene) were also then adopted to push towards single-defect operation. In the single-defect LED devices of Clark \textit{et al.}, the WSe$_2$ monolayer is sandwiched between two graphene electrodes and separated by thin hBN tunnel barriers, so that at zero bias the graphene Fermi levels lie in the WSe$_2$ bandgap, while under bias electrons (holes) tunnel from the negative (positive) electrode into sub-gap/near-gap states of the monolayer and rapidly bind into excitons due to strong Coulomb interactions \cite{clark2016single}. A practical outcome of this bipolar injection scheme is that EL turns on once the applied voltage is large enough to access these states. Clark \textit{et al.} report EL onset at $|V_b|\gtrsim 1.9~\mathrm{V}$ at $\sim 5$~K), with the spectrum near threshold dominated by defect-bound emission in the $\sim 1.65$--$1.71$~eV range, before intrinsic charged/neutral excitons appear at higher bias (peaks near $\sim 1.72$ and $\sim 1.75$~eV) \cite{clark2016single}. Because this vertical geometry injects carriers broadly across the active area, they further relied on spatial filtering to isolate individual hotspots, revealing that the EL from a single defect reproduces the characteristic fine-structure doublet and cross-linearly polarized selection rules familiar from optical single-photon-emitter studies \cite{clark2016single}.

Schwarz \textit{et al.} used closely related vdW stacks (monolayer WSe$_2$ with hBN tunneling barriers and graphene electrodes) to benchmark how barrier design and band alignment set both the injection window and the electrostatic control of defect emission. They also observed sharp defect spectra under electrical pumping and demonstrated vertical-field tuning of the defect luminescence by more than $1$~meV \cite{schwarz2016electrically}. At the single-emitter level, they reported narrow linewidths in the few-hundred-$\mu$eV regime (e.g., $245~\mu$eV for one defect at $V_b=-2.5$~V) together with a roughly linear Stark shift of order $0.4~\mathrm{meV/V}$ \cite{schwarz2016electrically}, and they emphasized that bias-dependent appearance/disappearance of lines and discontinuous spectral jumps are consistent with voltage-dependent charging in the local defect environment. Collectively, these studies demonstrated that vertical vdW stacks can behave as a quantum LED whose emission pathway is set by (i) the barrier-limited tunneling rates and (ii) the availability of localized sub-gap states that can efficiently capture injected carriers.

\begin{figure*}[t!] \centering
     \includegraphics[scale=0.7]{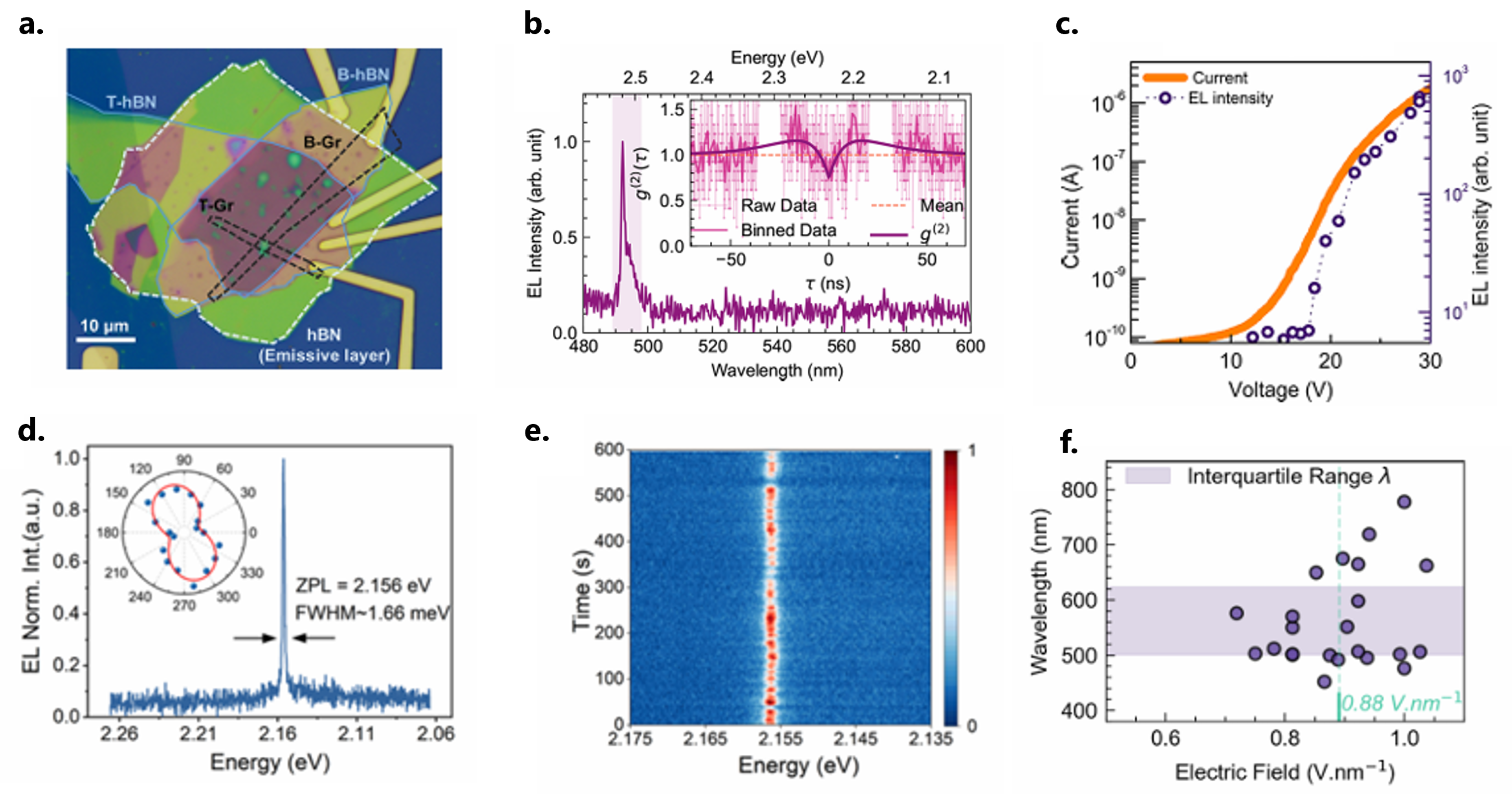}  
     \vspace{-0.5cm}
          \caption{\footnotesize \textbf{Electrical injection of quantum emitters in hBN.} \textbf{(a)} Optical micrograph of a graphene-contacted vdW heterostructure in which an emissive hBN layer is embedded in the vertical tunneling stack for electrically driven color-center EL (adapted from \cite{park2024narrowband}, Copyright 2024, American Chemical Society). \textbf{(b)} Representative narrowband EL spectrum from a color center under electrical pumping, with an inset showing the measured second-order autocorrelation $g^{(2)}(\tau)$ for electrically driven emission (adapted from \cite{park2024narrowband}, Copyright 2024, American Chemical Society). \textbf{(c)} Voltage-dependent current and EL intensity in the same device class, illustrating electrical turn-on behavior of a narrowband EL SQE (adapted from \cite{park2024narrowband}, Copyright 2024, American Chemical Society). \textbf{(d)} Low-background EL spectrum from a carbon-doped hBN device showing a dominant ZPL (with polarization inset) under electrical drive (adapted from \cite{wang2025low}, Copyright 2025, American Chemical Society). \textbf{(e)} Time-resolved spectral map of the ZPL intensity demonstrating spectral stability of electrically pumped emission over minute-scale operation (adapted from \cite{wang2025low}, Copyright 2025, American Chemical Society). \textbf{(f)} Electric-field threshold for in situ electrical generation of color centers in hBN, plotted as the distribution of SQE wavelengths versus applied field (adapted from \cite{zhigulin2025electrical}, Copyright 2025, American Chemical Society).}
          \label{fig: hBN EL}
\end{figure*}

While vertical QLEDs naturally provide area-wide injection, lateral devices provide a complementary route to spatially confined recombination by defining a narrow junction region in-plane. A representative implementation is the split-gate, monolayer p-i-n junction platform (Fig. \ref{fig: TMD EL}d) in which independent gate biases create adjacent p- and n-doped regions inside a single WSe$_2$ flake, localizing EL to the intrinsic junction \cite{lenferink2022tunable} with site-specific strain engineered SQE. Beyond compactness, the key advantage is that the junction electrostatics offer a direct tuning knob over the local charge environment of the SQE while maintaining electrical pumping. Experimentally, deterministic single-photon EL was demonstrated with strong antibunching for a monolayer emitter, $g^{(2)}(0)=0.085\pm 0.029$ (Fig. \ref{fig: TMD EL}e and \ref{fig: TMD EL}f), while a bilayer device yielded $g^{(2)}(0)=0.285\pm 0.034$ \cite{lenferink2022tunable}. These lateral junction devices also reported fine-structure splittings of $\sim 1.1$~meV (monolayer) and $\sim 0.5$~meV (bilayer), linking gate-defined recombination to excitonic anisotropy at the localized site. Notably, EL could be sustained at extremely low drive currents (e.g., $\sim 4.5$~nA in representative operating conditions), emphasizing the potential for low-power electrically driven quantum emission (Fig. \ref{fig: TMD EL}g. In this platform, the gate pair simultaneously (i) defines where electrons and holes overlap and (ii) tunes the local band bending experienced by the localized exciton, enabling spectral shifts exceeding 10~meV via electrostatic control of the junction region \cite{lenferink2022tunable}.

A conceptually distinct limit of electrical excitation is reached when injection is performed locally at the nanometer scale, enabling direct correlation between atomic-scale defects and photon emission. Schuler \textit{et al.} demonstrated electrically driven photon emission from individual atomic defects in monolayer WS$_2$, reporting photon emission yields on the order of $10^{-7}$ photons per injected electron \cite{schuler2020electrically}. While the injection mechanism and geometry differ from planar QLEDs, this work provides an important benchmark for defect-mediated radiative efficiency under purely electrical excitation at the single-defect level, and it highlights how strong localization and local fields can open radiative channels that are otherwise difficult to access in extended devices.

Recent efforts increasingly combine the robust injection of vertical vdW junctions with strategies that favor site-engineerable SQE formation and repeatable device operation. So \textit{et al.} (Fig.\ref{fig: TMD EL}c and \ref{fig: TMD EL}h) demonstrated a vertical junction approach that produces localized EL with low-voltage turn-on ($\sim 1.5$~V) and spectrally narrow lines (reported linewidths spanning $\sim 369$--$936~\mu$eV across representative SQEs) \cite{schuler2020electrically}. Under electrical pumping, they observed clear antibunching as low as $g^{(2)}(0)=0.13\pm 0.02$, alongside strong linear polarization (degree of polarization $\sim 0.946$--$0.965$) and fine-structure splitting of $\sim 755~\mu$eV, underscoring that electrically injected localized excitons at site-engineered locations can still preserve hallmark quantum-emitter signatures \cite{so2021electrically}. Guo \textit{et al.} then reported an electrically driven site-controlled single-photon source using gold pillars embedded in a graphene/hBN/WSe$_2$/hBN/graphene vertical tunneling LED \cite{guo2023electrically}. With thin hBN tunnel barriers, EL was localized to the pillar apex under bias (e.g., $V\approx 3.2$~V), and antibunching was quantified as $g^{(2)}(0)=0.32\pm 0.07$ \cite{guo2023electrically}. Crucially, by increasing the hBN barrier thickness to $>10$ layers, the same geometry could suppress tunneling-driven EL in favor of electric-field tuning under optical excitation. Howarth \textit{et al.} extended the vertical tunneling junction concept to SQE arrays and magneto-optical benchmarking, reporting electroluminescent quantum emitters spectrally $\sim 60$~meV below the neutral exciton and extracting excitonic $g$-factors spanning approximately $-4$ to $-15$ \cite{howarth2024electroluminescent}. Their analysis further emphasized that the relevant spin/valley structure is set by the large intrinsic band splittings (representatively $\sim 30$~meV in the conduction band and $\sim 510$~meV in the valence band), motivating vertical junction designs that can selectively populate and read out localized excitonic states under electrical drive.

Taken together, these works trace the evolution of electrically driven TMD quantum emitters from early vertical vdW QLEDs demonstrating antibunched EL in monolayer WSe$_2$ ($g^{(2)}(0)\approx 0.29$ at 4.2~K) \cite{palacios2018atomically}, through improved understanding and control of defect-bound and single-defect injection \cite{clark2016single,schwarz2016electrically}, to devices that localize recombination electrostatically (split-gate p-i-n junctions reaching $g^{(2)}(0)\sim 0.085$ at nA currents) \cite{lenferink2022tunable}, and toward scalable vertical-junction implementations that benchmark sub-meV fine-structure, few-hundred-$\mu$eV linewidths, low turn-on voltages ($\sim 1.5$~V), and wide electrical tunability (up to $\sim 40$~meV) \cite{so2021electrically,guo2023electrically,howarth2024electroluminescent}.

\subsubsection{Hexagonal Boron Nitride}

We next turn to electrically driven emission from defect centers in hexagonal boron nitride (hBN), where wide-bandgap defect physics and defect-state-assisted transport motivate related but distinct device designs for EL operation. Electrical injection into hBN color centers is governed by different microscopic physics than in TMD-based QLEDs as hBN is a wide-bandgap semiconductor, so transport and excitation are often mediated by defect-state-assisted tunneling and high-field processes rather than conventional band-to-band injection. Nevertheless, the benchmarking metrics are largely the same as for electrically driven TMD emitters such as (i) EL turn-on bias/current and power consumption, (ii) spectral purity and linewidth, (iii) single-photon purity via $g^{(2)}(0)$, (iv) spectral stability (blinking/spectral diffusion over minutes to hours), and (v) electrical control knobs (tuning ranges and reproducibility). The main difference is therefore not what is benchmarked, but how device design and defect physics set those benchmarks in an insulating host. Figure \ref{fig: hBN EL} highlights three representative thread focusing on narrowband EL from color centers in graphene/hBN tunnel junctions (Fig. \ref{fig: hBN EL}a-c), low-background single-photon EL from engineered carbon-doped hBN (Fig. \ref{fig: hBN EL}d and e), and electrically generating color centers via high-field operation (Fig. \ref{fig: hBN EL}f).

\begin{figure*}[t!] \centering
     \includegraphics[scale=0.65]{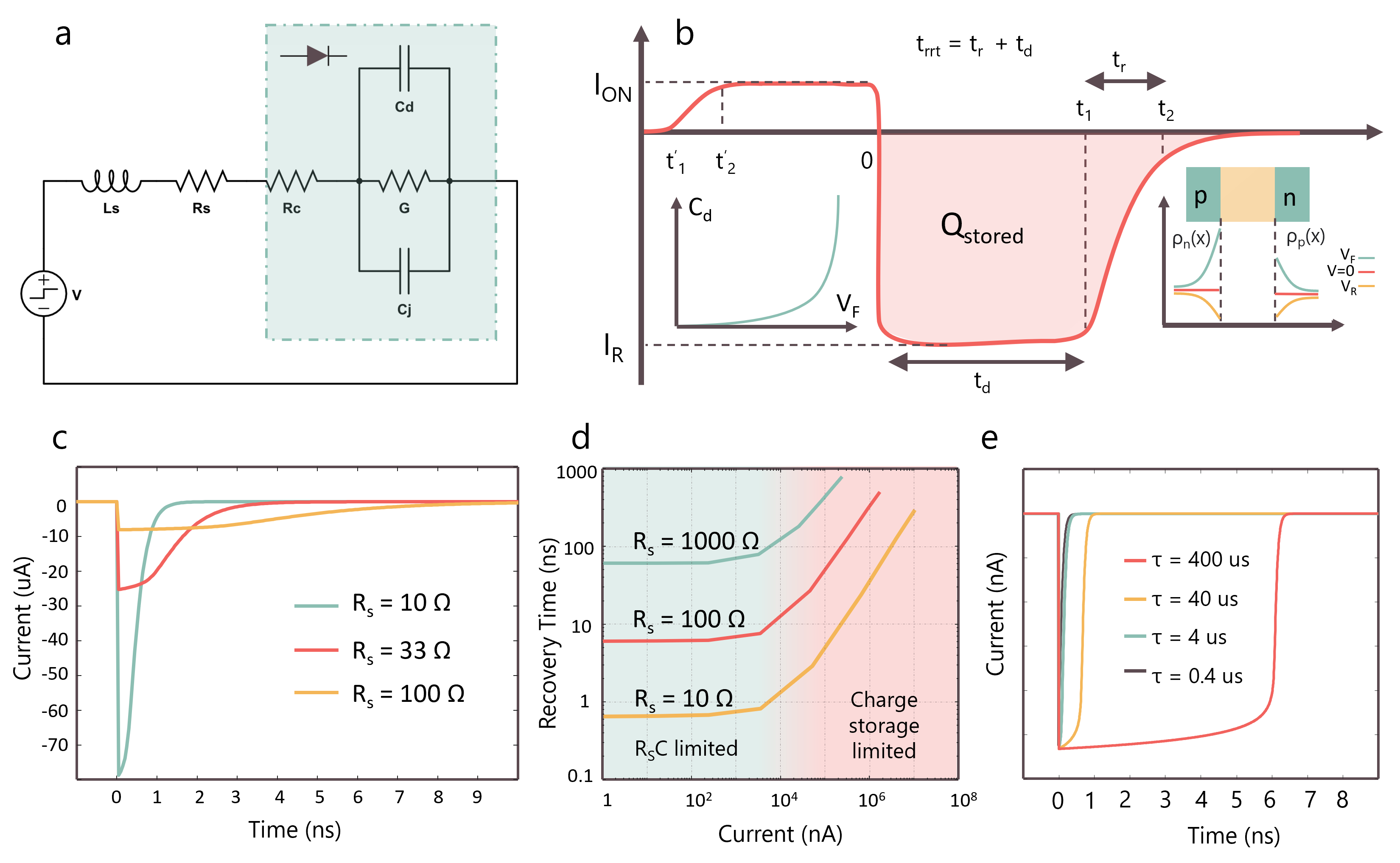}  
     \vspace{-0.5cm}
          \caption{\footnotesize \textbf{Switching behaviour of PN-based quantum LEDs.} \textbf{(a)} Small signal model of a PN diode. $L_s$, $R_s$, $R_c$, $C_d$, $C_j$, and $G$ denote the series inductance, series resistance, contact resistance, diffusion capacitance, junction capacitance, and conductance, respectively. \textbf{(b)} Transient response of PN diodes in response to a pulsed signal. \textbf{(c)} Effect of contact resistance on the diode. \textbf{(d,e)} Effect of current and carrier lifetime on reverse recovery time, respectively.}
          \label{fig: QLED}
\end{figure*}

A central demonstration for electrically driven hBN SQE by Park \textit{et al.} is that a vdW tunneling stack can produce EL dominated by discrete ZPL-like features rather than a broad background. In this narrowband EL study \cite{park2024narrowband}, the device is a vertical graphene/hBN/graphene junction in which the emissive hBN layer is embedded within the hBN tunneling barrier and addressed by top and bottom graphene electrodes (Fig. \ref{fig: hBN EL}a). Under applied bias, defect levels within hBN enter the transport window and mediate current through resonant or inelastic tunneling pathways where radiative decay from these defect states then yields spectrally sharp EL. Experimentally, EL spectra from SQEs were reported at cryogenic temperature ($\sim 10$~K) with ZPLs spanning the visible spectrum. A practical benchmark from these devices is their comparatively high turn-on bias for the \(\sim 490\)~nm SQE where the EL intensity was observed to rise steeply from a voltage on the order of \(\sim 18\)~V (Fig. \ref{fig: hBN EL}c). Consistent with the localized-defect picture, Park \textit{et al.} show that the emission can be strongly concentrated into a narrow spectral window. For example, representative narrowband EL features were discussed with linewidths on the order of a few nanometers (e.g., \(\sim 5\)~nm, corresponding to \(\sim 25\)~meV in the visible). Second-order correlation measurements were also reported in this work (Fig. \ref{fig: hBN EL}b inset), with a representative value of \(g^{(2)}(0)=0.77\), highlighting that while narrowband EL is readily achieved, the single-photon purity can be limited by background and/or multi-center contributions unless the emitting site and injection pathway are well isolated. This study established the spectral selectivity of electrically pumped hBN color centers and clarified that the relevant device knob is the bias-controlled alignment of graphene chemical potentials with defect manifolds in the hBN barrier \cite{park2024narrowband}.

A complementary route is to engineer the emissive defect population directly in the electrically active layer. In the carbon-doped hBN study by Wang \textit{et al.} \cite{wang2025low}, the key result is low-background EL where individual ZPLs dominate and antibunching is observed under electrical pumping. Figure \ref{fig: hBN EL}d shows a representative spectrum from a ZPL at \(E_{\mathrm{ZPL}} \approx 2.156\)~eV with a reported linewidth \( \mathrm{FWHM} \approx 1.66\)~meV, together with strong linear polarization (Fig. \ref{fig: hBN EL}d inset). Wang \textit{et al.} also demonstrated antibunching from the electrically driven ZPL with \(g^{(2)}(0)\approx 0.4\), directly evidencing single-photon EL with comparatively low spectral background \cite{wang2025low}. Beyond single-photon purity, this platform emphasized spectral stability where Fig. \ref{fig: hBN EL}e shows a time-resolved spectral map of the ZPL over \(\sim 600\)~s, and the study reports stability measurements extending to \(\sim 1200\)~s for representative SQEs \cite{wang2025low}. Taken together, these benchmarks position carbon-doped hBN as an attractive electrically driven defect platform where the active barrier itself is engineered to favor narrowband, low-background EL.

A third study emphasizes that electrical bias can do more than excite pre-existing defects---it can create new optically active centers in situ \cite{zhigulin2025electrical}. In this study, Zhigulin \textit{et al.} showed that the key benchmarking metric is the electric-field threshold required to activate new color-center EL. This work reports that devices typically required a mean electric field of \(\sim 0.88~\mathrm{V\,nm^{-1}}\) to generate color centers, after which new narrow peaks emerged in the EL spectrum (Fig. \ref{fig: hBN EL}f) \cite{zhigulin2025electrical} . A representative temporal benchmark is that, once the device is operated in this high-field regime, a distinct color-center peak can appear after \(\sim 15\)~minutes (with an example peak reported near \(\sim 390\)~nm), followed by evolution in intensity over time. From a device-physics perspective, these observations are consistent with field-driven defect formation, charge-state rearrangement, or local chemical modification under tunneling/field stress processes. Practically, this introduces an additional ``integration knob'' unique to hBN EL where electrical operation can both write and read defect SQEs, enabling device-defined activation protocols at the cost of requiring high fields and careful management of device degradation.

Electrically pumped single-photon emission has also been reported in related vdW heterostructures where the tunneling current is deliberately confined to a small active region to better isolate individual emitting sites \cite{yu2024electrically,grzeszczyk2024electroluminescence}. In one such platform, background-corrected photon statistics yielded $g^{(2)}(0)=0.25\pm 0.21$, and representative stability measurements were acquired at biases on the order of $\sim 30$~V \cite{yu2024electrically}. More broadly, these results complement the narrowband tunnel-junction demonstrations and engineered carbon-doped hBN SQEs \cite{park2024narrowband,wang2025low}, while also aligning with the observation that electrical operation can modify or activate defect populations under sufficiently strong fields \cite{zhigulin2025electrical}. Together, these works reinforce that quantum-grade hBN EL hinges on managing the competition between radiative recombination within a defect manifold and non-radiative tunneling escape and/or field-induced charge-state dynamics through the barrier \cite{park2024narrowband,wang2025low,grzeszczyk2024electroluminescence,zhigulin2025electrical,yu2024electrically}. Therefore, the dominant benchmarks for progress are (i) lowering EL turn-on toward the nA--$\mu$A tunneling regime, (ii) maximizing ZPL dominance and minimizing broadband background, (iii) improving single-photon purity, (iv) extending spectral stability, and (v) developing deterministic activation/tuning protocols. These metrics set the stage for comparing hBN EL devices directly with TMD-based electrically driven SQEs, while highlighting the distinct physical constraints and opportunities of defect-state electroluminescence in an wide-bandgap semiconducting 2D host.

\subsection{Towards High-Rate Pulsed Operation}

While many 2DM QLED demonstrations operate in a DC or slowly modulated regime, quantum-photonic systems ultimately require pulsed operation with a well-defined clock where single photons are emitted in discrete time bins, triggered on demand, and at MHz to GHz rates is therefore not merely an engineering refinement but the operating point where electrically pumped 2D emitters begin to resemble practical quantum-light engines. At these rates, the fundamental optical lifetime sets an approximate upper bound, but the dominant constraints are typically electrical parasitic impedances in the device stack and wiring, carrier storage in the diode under forward bias, and the time required to return the junction to a true ``off'' state between pulses. Accordingly, the key question then becomes focused on parameters that set the electrical switching time of a nanoscale 2D PIN junction, and which knobs most directly move it toward the GHz regime

Figure~\ref{fig: QLED} summarizes the relevant high-speed physics using the standard small-signal description of a PN junction \cite{ng2007physics}. In Fig.~\ref{fig: QLED}a, the junction is modeled by a bias-dependent conductance $G$ in parallel with a junction capacitance $C_j$ (depletion-region charging) and a diffusion capacitance $C_d$ (minority-carrier storage during forward bias), together with parasitics $L_s$, $R_s$, and---especially important in 2D monolayer devices, contact resistance $R_c$. This circuit picture connects directly to the transient waveform in Fig.~\ref{fig: QLED}b where as the drive switches from reverse bias (OFF) to forward bias (ON), the turn-on can be fast if inductive elements are minimized. However, the most severe bottleneck is often the turn-off. After the voltage crosses zero, the diode can remain conductive because minority carriers stored on either side of the depletion region must be removed. This produces a reverse current transient and a reverse-recovery time $t_{rr}=t_d+t_t$, which ultimately limits how quickly successive pulses can be applied without temporal overlap.

This framework reveals three device-level levers. First, lowering series resistance (in practice, dominated by contacts and access regions in 2D PN junctions) reduces the RC-limited rise and fall times where Fig.~\ref{fig: QLED}c illustrates how increasing $R_s$ slows the current transient, motivating contact engineering and impedance-matched interconnect design \cite{ng2007physics,schulman2018contact,prakash2017understanding}. Second, minimizing diffusion capacitance by operating at the lowest feasible ON current reduces stored charge and shortens recovery where Fig.~\ref{fig: QLED}d highlights the crossover between an $R_cC$-limited regime at low current and a charge-storage-limited regime at higher current, emphasizing the value of high internal quantum efficiency that enables EL at nA--$\mu$A currents \cite{ng2007physics,lenferink2022tunable}. Third, reverse recovery scales with the minority-carrier lifetime, so faster recombination shortens $t_{rr}$ where Fig.~\ref{fig: QLED}e shows the reduction in recovery with decreasing carrier lifetime (or equivalently faster recombination), motivating lifetime engineering through controlled defect densities or radiative-rate enhancement (e.g., Purcell enhancement) where compatible with emitter quality \cite{ng2007physics,wang2015ultrafast,wu2016defects}.

In practice, these levers translate into concrete integration directions towards GHz-class 2D QLEDs. Contact resistance can be reduced using contact gating with overlapping local gates with the metal--semiconductor region to narrow Schottky barriers and by selecting low-barrier, minimally pinned contact schemes, including high-quality metal deposition, transferred/clean interfaces, and vdW or semimetal contacts where appropriate \cite{schulman2018contact,prakash2017understanding,jung2019transferred,english2016improved,wang2019van,shen2021ultralow,li2023approaching}. At the circuit level, short/wide leads and tapered traces mitigate parasitic $L_s$ and $C$ while improving impedance matching for fast edges, as demonstrated in cryogenic RF delivery to 2D devices \cite{patel2024surface,zhang2019two,yang2020ultrafast}. As the charge-storage constraint relaxes rapidly as the required injection current decreases, improvements in emitter quantum yield (and, where applicable, cavity-enhanced radiative rates) directly ease the GHz requirement by allowing operation deep in the low-current, low-$C_d$ regime \cite{lenferink2022tunable}. Together, Fig.~\ref{fig: QLED}a-e provides a compact design rule for high-rate towards GHz pulsed operation is enabled when contact/series RC, diffusion charge storage, and carrier lifetime are co-optimized so that electrical transients no longer dominate the optical repetition rate.

\subsection{Stark Tuning and Charge Stabilization}

\begin{figure*}[t!] \centering
     \includegraphics[scale=1]{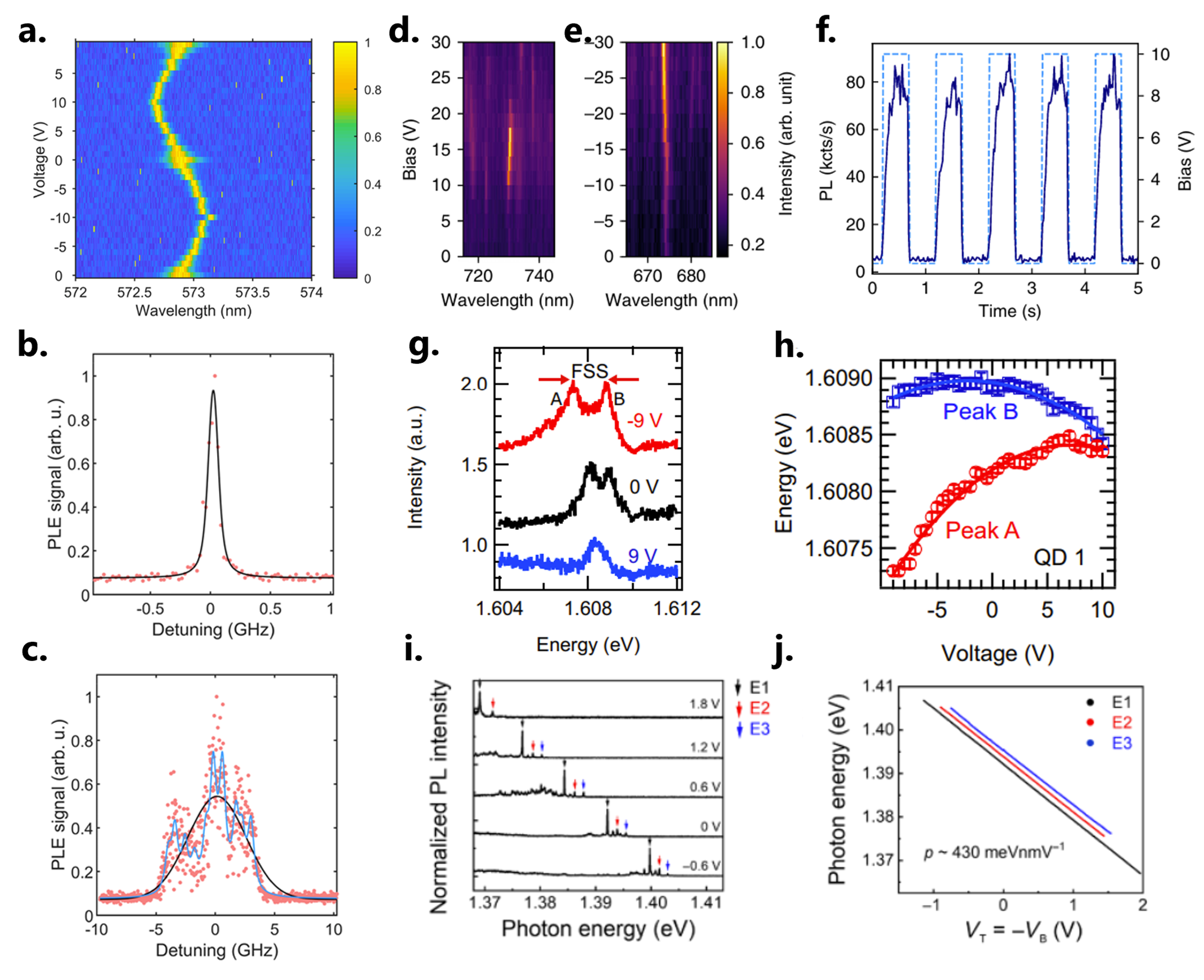}  
     \vspace{0 cm}
          \caption{\footnotesize \textbf{Electrical tunability of quantum emitters in hBN and TMDs.} \textbf{(a)} Representative stark-tuning map with field-dependent narrowing and stabilization of an hBN ZPL enabled by electric-field modulation in a gated device (adapted from \cite{akbari2022lifetime}, Copyright 2022, American Chemical Society). \textbf{(b)} Resonant excitation under electrical control highlighting coherence-grade operation in hBN with representative near-transform limited linewidth of 89 MHz (adapted from \cite{akbari2022lifetime}, Copyright 2022, American Chemical Society). \textbf{(c)} Resonant excitation under no electrical control highlighting broadened hBN ZPL linewidth with spectral diffusion (adapted from \cite{akbari2022lifetime}, Copyright 2022, American Chemical Society). \textbf{(d)} Bias-dependent spectral map showing electrically addressable emitter lines in a vdW heterostructure under gating (adapted from \cite{white2022electrical} under the CC BY 4.0 license). \textbf{(e)} Voltage-controlled spectral selection and switching of emission features in the same platform (adapted from \cite{white2022electrical} under the CC BY 4.0 license). \textbf{(f)} Dynamic voltage modulation demonstrating electrically driven on/off control of the optical output (adapted from \cite{white2022electrical} under the CC BY 4.0 license). \textbf{(g)} Electrical manipulation of the FSS of a WSe$_2$ quantum emitter via gate-controlled fields (adapted from \cite{chakraborty2019electrical}, Copyright 2019, American Physical Society). \textbf{(h)} Gate-dependent tuning of the FSS and emission energy in the same device class (adapted from \cite{chakraborty2019electrical}, Copyright 2019, American Physical Society). \textbf{(i)} Large DC Stark tuning of moir\'e-trapped interlayer excitons, showing highly energy-tunable quantum emission (adapted from \cite{baek2020highly} under the CC BY 4.0 license). \textbf{(j)} Stark-tuning and field-tunable emission characteristics of moir\'e-trapped excitons (adapted from \cite{baek2020highly} under the CC BY 4.0 license).}
          \label{fig: electrical tuning}
\end{figure*}

A central promise of electrical addressability is not only the ability to turn SQEs on, but also the ability to tune and stabilize them. Across both TMD and hBN platforms, gate electrodes provide a direct knob over the local electrostatic environment where an applied out-of-plane field can shift optical transition energies via the quantum-confined Stark effect (QCSE), while control of carrier density and trap occupation can suppress charge-noise-driven spectral wandering, mitigate blinking, and stabilize a desired charge state. In practice, these two functions of spectral tunability and charge stabilization are deeply coupled, because the same nearby charges that enable large tuning can also broaden lines if they fluctuate.

Early demonstrations established that hBN color centers can be tuned substantially by an external field, with behavior consistent with a combination of linear (permanent dipole) and quadratic (polarizability) Stark responses \cite{noh2018stark,scavuzzo2019electrically,nikolay2019very,mendelson2019engineering,xia2019room,zhigulin2023stark}.
In a representative capacitor-style geometry, Noh \textit{et al.} reported Stark shifts as large as $\sim$5.4~nm per GV/m and extracted defect-scale parameters ($\Delta \mu \approx -0.22$~D and $\Delta \alpha \approx 149~\text{\AA}^3$), while maintaining antibunching with $g^{(2)}(0)\approx 0.3$ at $\sim$10~K \cite{noh2018stark}. Complementary device concepts leveraged vdW stacks with graphene serving as an ultrathin gate/contact. For example, Scavuzzo \textit{et al.} observed linear Stark tuning up to $\sim$20~meV with slopes approaching $\sim$24~meV/(V/nm), and emphasized robustness over repeated gate cycles (with occasional spectral jumps of several meV superposed on the deterministic Stark response) \cite{scavuzzo2019electrically}. At room temperature, Nikolay \textit{et al.} demonstrated reversible tuning of $\sim$5.5 $\pm$ 0.3~nm (15.4 $\pm$ 0.8~meV) at $\lambda\approx 670$~nm using $\sim$20~V bias, underscoring that sizeable quantum-confined Stark effect (QCSE) persists beyond cryogenic operation \cite{nikolay2019very}. Beyond conventional capacitive device structures, ionic-liquid gating in few-layer hBN enabled particularly wide tuning windows where Mendelson \textit{et al.} reported ZPL tuning over \,$>$15~nm at room temperature, while preserving single-photon character across many SQEs \cite{mendelson2019engineering}. At the high-field extreme, large Stark behavior has also been reported, including $\sim$31~meV shifts at fields $\sim$0.36~V/nm (corresponding to an effective tuning coefficient of $\sim$43~meV/(V/nm)) \cite{xia2019room}. At the same time, not all centers are equally field-sensitive where for example, blue-emitting centers can show much weaker (often predominantly quadratic) responses, with reported sensitivities on the order of $\sim$0.04~meV~V$^{-1}$~nm \cite{zhigulin2023stark}.

Against this backdrop, recent work has increasingly emphasized low-noise and high optical coherence operation under electrical control with simultaneous achievement of tunability and narrow optical linewidths \cite{akbari2022lifetime,white2022electrical, paralikis2025tunable, hotger2021gate, chen2023gate}. Figure \ref{fig: electrical tuning} highlights this direction where in Fig.~\ref{fig: electrical tuning}a-c, lifetime-limited and tunable hBN emission is enabled by electric-field modulation in a gated heterostructure, where linewidths can be driven down to the $\sim$100~MHz scale with reported ZPL linewidths as low as 89 MHz under resonant excitation while retaining deterministic field tuning \cite{akbari2022lifetime}. In the same study, the measured Stark tunability reaches $\sim$2.7~meV/(V/nm), providing a quantitative bridge between device-level tuning and coherence metrics \cite{akbari2022lifetime}. (Fig.~\ref{fig: electrical tuning}d--f) further emphasize that electrically controlled heterostructures can be operated dynamically. White \textit{et al.} used a graphene-gated vdW stack to tune and select defect emission in real time, visualized directly in bias-dependent spectral maps (Fig.~\ref{fig: electrical tuning}d,e) where discrete ZPL features are brought in and out of resonance as the applied bias changes. Furthermore, they showed electrical switching of the optical output under a square-wave bias (Fig. \ref{fig: electrical tuning}f), demonstrating repeatable on/off modulation on $\sim$sub-second time scales. In the same work, resonant excitation under an optimized vertical field yielded near-transform-limited linewidths as low as $158\pm 19$~MHz, while second-order correlation measurements remained consistent with single-photon emission, highlighting that electrical gating can simultaneously provide a tuning/stabilization knob and preserve quantum emission characteristics \cite{white2022electrical}. 

Related device studies likewise highlight that electrical control is inseparable from defect charging where Yu \textit{et al.} quantify Stark slopes spanning $\sim$0.029--0.86~meV/(MV/cm) across SQEs (with an extracted dipole on the order of $\sim$0.09~D for representative cases) and demonstrate stability over thousands of voltage cycles \cite{yu2022electrical}. More recently, current-driven reconfiguration has been used as an in situ knob to brighten vacancy-related centers which demonstrates how injection and gating can reshape the local charge landscape even when the primary goal is optical stabilization rather than large tuning \cite{steiner2025current}. The push toward deterministic charge configuration extends naturally to spin-active defects where charge-state tuning of spin defects in hBN provides an emerging pathway to stabilizing optically addressable spin manifolds under electrical control \cite{fraunie2025charge}.

In monolayer and heterobilayer TMD SQEs, electrical control has also been pursued both as (i) a direct QCSE knob on localized excitons/defects and (ii) a route to suppress charge noise and improve quantum efficiency. An early, clear demonstration of QCSE in localized WSe$_2$ emission also used vertical heterostructure capacitors to apply strong out-of-plane fields to individual defect-like SQEs, achieving efficient spectral tunability up to $\sim$21~meV \cite{chakraborty2017quantum}. A complementary route uses field control to interrogate polarizability itself where Mukherjee \textit{et al.} reported strain-induced SQEs in nanostructured WSe$_2$ that can exhibit anomalous blue shifts under field, consistent with negative polarizability and strong device-to-device variability spanning orders of magnitude, highlighting that tuning and noise share a common microscopic origin in local charge configuration \cite{mukherjee2020electric}.

A particularly instructive example of deliberate charge engineering is provided by Cai \textit{et al.}, who combine nanogap-enabled localization with dual-gate electrostatics to suppress charge-assisted loss channels in WSe$_2$ single-photon emitters (SPEs) \cite{cai2024charge}. In this platform, the emission is strongly gate dependent: mapping the optical response versus top- and back-gate voltages reveals distinct electrostatic operating regions with pronounced intensity modulation, showing that the SPE output is governed by the combined gate configuration rather than by a single bias alone \cite{cai2024charge}. Time-resolved measurements further show that the lifetime varies in close correspondence with the intensity across the same gate space, directly implicating a gate-modulated non-radiative decay rate as the dominant electrical knob \cite{cai2024charge}. The authors attribute this behavior to vertical-field-induced redistribution of the surrounding charge environment toward a dielectric interface, which reduces spatial overlap between nearby charges and the neutral localized exciton, thereby suppressing non-radiative pathways such as trion formation and Auger-assisted recombination \cite{cai2024charge}. Consistent with this depletion picture, they report substantially improved single-photon performance and extract large increases in transition quantum efficiency under gating; across a statistical set, the post-gating average reaches $\eta_q=76.4\pm14.6\%$, with some SPEs exceeding $90\%$ \cite{cai2024charge}.

Electrical control also enables manipulation of internal excitonic fine structure. In Fig. \ref{fig: electrical tuning}g-h, Chakraborty \textit{et al.} use gate-controlled fields to tune the exciton doublet splitting in WSe$_2$ quantum emitters, demonstrating fine-structure modulation up to $\sim$1500~$\mu$eV with reported zero-field splittings up to $\sim$800~$\mu$eV, and Stark shifts on the order of hundreds of $\mu$eV across the tuning range \cite{chakraborty2019electrical}. Beyond monolayers, moir\'e heterobilayers provide a strikingly large and highly linear tuning knob as shown in Fig. \ref{fig: electrical tuning}i-j where Baek \textit{et al.} achieve DC Stark tuning up to $\sim$40~meV for moir\'e-trapped interlayer excitons and extract a large permanent dipole moment of $p \approx 429\pm 4$~meV\,nm\,V$^{-1}$, while maintaining single-photon statistics with $g^{(2)}(0)=0.28\pm 0.03$ \cite{baek2020highly}. The field has also expanded the materials and wavelength space of electrically controlled SQEs where recent telecom MoTe$_2$ quantum emitters show that electrostatic biasing can provide active Stark tuning over $\sim$3~meV while enabling high purity ($g^{(2)}(0)<0.1$) and measurable two-photon interference (HOM visibility $\sim$10\%, increasing up to $\sim$40\% with temporal post-selection), directly tying charge stabilization and tuning to indistinguishability metrics \cite{wyborski2025toward}.

It is worth emphasizing that electrical tunability need not be purely electrostatic where electro-mechanical control can be equally powerful. Suspended WSe$_2$ devices driven by electrostatically induced strain demonstrate dynamic spectral control at the tens-of-meV level, accompanied by linewidth narrowing \cite{wu2025modulation}. Related strain-based approaches can generate SQEs at high spatial density (e.g., $\sim$150 emitters/$\mu$m$^2$ in nanoparticle-array-induced strain landscapes), motivating hybrid electrical--strain strategies where gates stabilize charge while strain pins localization \cite{kim2022high,chakraborty2015voltage}. Across TMD and hBN SQEs, the experimental toolkit for Stark tuning and charge control has matured into a coherent set of device concepts where capacitor-like vdW stacks and graphene-gated heterostructures provide large QCSE tunability (tens of meV in favorable cases), while carefully engineered electrostatics (depletion, gating, and controlled charging) increasingly sets the benchmark for single-photon purity, linewidth, stability, and---in emerging telecom-band TMD SQEs even photon indistinguishability.

\section*{Photonic Integration}

Photonic integration provides a scalable route to interface 2D material quantum SQEs with low-loss optical circuitry for quantum communication and on-chip quantum optics. By coupling emission into engineered photonic modes, integrated platforms can dramatically improve photon collection and directionality, enable deterministic routing and multiplexing, and provide a natural pathway to interfere photons from remote SQEs on a common circuit. At the same time, nanophotonic environments offer additional functionality beyond collection, including spectral and polarization control, Purcell enhancement, and access to chip-level filtering and switching. The following sections survey three complementary integration strategies, progressing from broadband waveguide integration, to enhanced light--matter interaction in on-chip cavities, and to high-directionality, free-space interfaces enabled by off-chip cavities.

\subsection{Waveguide Integration Methods}

\subsubsection{Transition Metal Dichalcogenides}

\begin{figure*}[t!] \centering
     \includegraphics[scale=0.6]{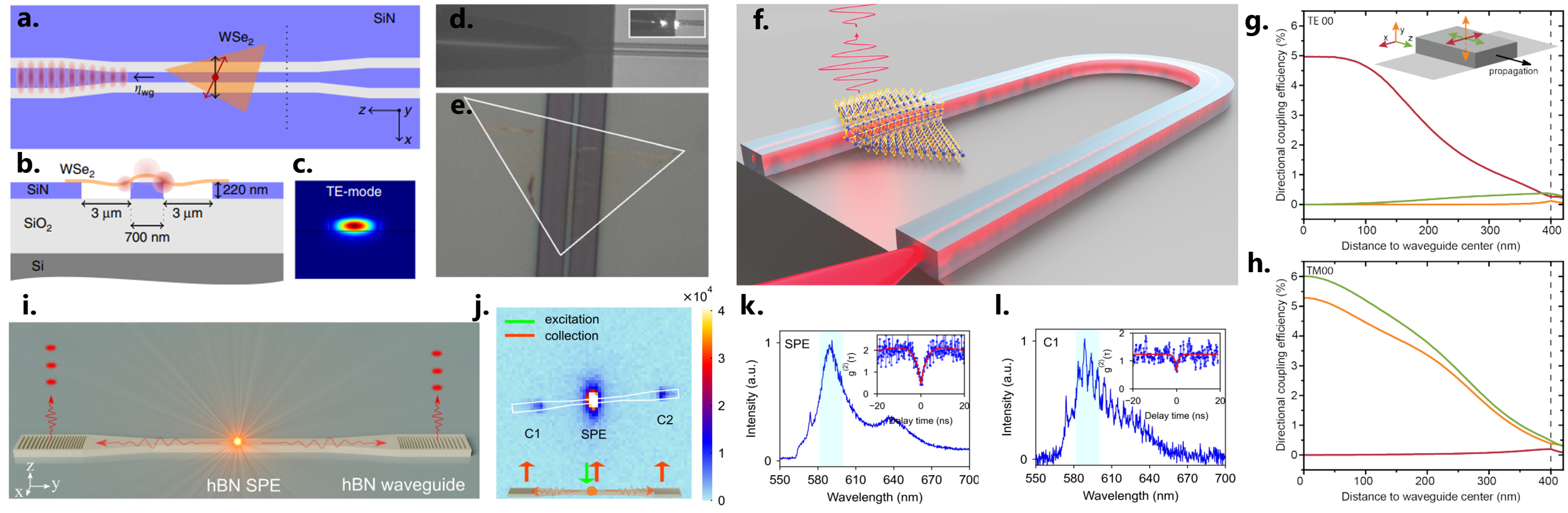}  
     \vspace{-0.5cm}
          \caption{\footnotesize \textbf{Photonic waveguide integration approaches.} \textbf{(a)} Conceptual schematic of a TMD monolayer transferred onto a SiN waveguide, where strain-localized exciton SQEs couple evanescently into the guided mode for in-plane routing (adapted from \cite{peyskens2019integration} under the CC BY 4.0 license). \textbf{(b)} Cross-sectional illustration of the non-embedded integration geometry, where the monolayer sits on the top surface of the waveguide, and waveguide edges generate a localized strain field for SQE creation while the guided mode provides a near-surface electric field for partial capture of the SQE's emission (adapted from \cite{peyskens2019integration} under the CC BY 4.0 license). \textbf{(c)} Simulated fundamental TE mode intensity profile of the waveguide, highlighting the evanescent field overlap available to TMD SQEs (adapted from \cite{peyskens2019integration} under the CC BY 4.0 license). \textbf{(d)} Optical micrograph of the prefabricated waveguide platform, showing the region used for integrating the transferred monolayer and accessing waveguide-coupled emission (adapted from \cite{peyskens2019integration} under the CC BY 4.0 license). \textbf{(e)} Optical micrograph of a transferred TMD flake positioned over the waveguide (adapted from \cite{peyskens2019integration} under the CC BY 4.0 license). \textbf{(f)} Illustration of a non-embedded TMD monolayer integrated on a SiN waveguide, where strain activation along the waveguide edges generates localized quantum emitters that evanescently couple a fraction of their emission into the guided mode (adapted from \cite{errando2021resonance} under the CC BY 4.0 license). \textbf{(g)} 3D-FDTD simulation of the directional coupling efficiency into the fundamental quasi-TE mode as a function of the SQE's lateral displacement from the waveguide center, highlighting strong sensitivity to dipole orientation, misalignment, and position offset (adapted from \cite{errando2021resonance} under the CC BY 4.0 license). \textbf{(h)} 3D-FDTD simulation of the directional coupling efficiency into the fundamental quasi-TM mode versus SQE position across the waveguide, emphasizing the limited coupling expected for edge-localized strain-activated SQEs in the non-embedded geometry (adapted from \cite{errando2021resonance} under the CC BY 4.0 license). \textbf{(i)} Schematic of a monolithic hBN waveguide platform hosting an embedded hBN SPE, with emission guided in-plane to out-of-plane grating couplers at both ends for on-chip routing and collection (adapted from \cite{li2021integration}, Copyright 2021, American Chemical Society). \textbf{(j)} Confocal intensity map of the device under local excitation, showing a bright SPE within the waveguide and simultaneous signal at the two out-couplers, confirming guided emission from the SQE to the chip outputs (adapted from \cite{li2021integration}, Copyright 2021, American Chemical Society). \textbf{(k)} PL spectrum collected at the SPE location, with the inset showcasing a second-order autocorrelation measurement verifying single-photon emission (adapted from \cite{li2021integration}, Copyright 2021, American Chemical Society). \textbf{(l)} PL spectrum collected at the output coupler, demonstrating that the same SQE emission is routed within the waveguide to the output, with the inset highlighting the preservation of the single-photon statistics after waveguide propagation and out-coupling (adapted from \cite{li2021integration}, Copyright 2021, American Chemical Society).}
          \label{fig: waveguide integration}
\end{figure*}

A dominant theme in TMD waveguide integration is the non-embedded or ``surface-mounted'' strategy, in which a monolayer (most commonly WSe$_2$) is dry-transferred onto a pre-fabricated dielectric nano-waveguide (typically SiN). In this geometry, localized strain gradients introduced by waveguide topography and edges, as well as transfer-induced wrinkles, cracks, and local nonuniformities, can seed narrow-linewidth SQEs directly in the monolayer while simultaneously positioning those SQEs in the near field of the guided mode (Fig. \ref{fig: waveguide integration}a--e). In this surface-mounted configuration, the SQE is not embedded in the waveguide core, so coupling to the photonic integrated circuit occurs through evanescent overlap with the fundamental guided mode (often quasi-TE), which robustly yields partial waveguide capture whenever the SQE lies sufficiently close to the optical mode. However, the waveguide-coupled fraction is typically highly emitter-dependent, because it depends on both the SQE's spatial offset from the mode maximum and the projection of the (generally uncontrolled) dipole orientation onto the guided-mode polarization. As a result, a bare waveguide can provide reliable on-chip routing but generally cannot drive near-unity extraction into a single guided mode, motivating later ``funneling'' concepts (e.g., cavities) when high $\beta$-factors are required.

These ideas are illustrated by an archetypal demonstration by Peyskens \textit{et al.}. \cite{peyskens2019integration}, which shows that integrating WSe$_2$ on SiN produces strain-associated emitters near waveguide and crack regions and that these SQEs can be evanescently coupled into the waveguide (Fig. \ref{fig: waveguide integration}a--e). A closely related implementation is shown by Errando-Herranz \textit{et al.}. \cite{errando2021resonance}, who again use waveguide-edge strain activation after dry transfer and emphasize that SQE creation and waveguide proximity can occur in the same scalable step (Fig. \ref{fig: waveguide integration}f). Importantly, this work makes the role of dipole and placement disorder explicit through 3D-FDTD sweeps of lateral dipole position across the waveguide width and dipole orientation, mapping these imperfections onto the expected coupling into the fundamental quasi-TE (TE$_{00}$) and quasi-TM (TM$_{00}$) modes (Fig. \ref{fig: waveguide integration}g, h). In the edge-localized configuration that is natural for strain activation, the reported orientation-averaged unidirectional coupling efficiency into the fundamental modes is limited to approximately 0.32\% (TE$_{00}$) and 0.34\% (TM$_{00}$) \cite{errando2021resonance}, illustrating a practical ceiling for plain edge-strain coupling without additional photonic engineering.

Beyond SiN, similar ``transfer-onto-existing-photonics'' logic has also been used to interface TMD SQEs with other low-loss waveguide platforms. For example, White \textit{et al.} \cite{white2019atomically}, transfer monolayer WSe$_2$ onto a Ti in-diffused lithium niobate directional coupler, highlighting a route to combine atomically thin emitters with a mature, electro-optically active integrated photonics material system. Related on-chip waveguide coupling has also been demonstrated for other layered semiconductors, for example by Tonndorf \textit{et al.} \cite{tonndorf2017chip}, in a GaSe-on-SiN platform, underscoring the generality of transfer-based near-field coupling for 2D SQEs beyond TMDs.

Overall, these non-embedded demonstrations provide a scalable and fabrication-light methodology to route quantum emission in-plane on chip, while also making clear that deterministic strain placement, dipole control, and photonic funneling will be central to pushing waveguide extraction toward high-$\beta$ and device-uniform operation.

\subsubsection{Hexagonal Boron Nitride}

In contrast to the largely transfer-onto-existing-photonics approaches used for TMDs, most hBN waveguide demonstrations have favored monolithic integration, where photonics are patterned directly into the hBN host itself. Figure \ref{fig: waveguide integration}i--l highlights a particularly popular and conceptually simple strategy for hBN waveguide integration, namely monolithic hBN nanophotonics, where the host material and the guiding structure are the same. In this approach, the waveguide is patterned directly from an exfoliated hBN flake, and SQEs are then created after photonic fabrication through post-treatment, after which randomly occurring SPEs that happen to fall within the waveguide region are identified \cite{li2021integration}. The specific device illustrated here (Fig. \ref{fig: waveguide integration}i) uses a slot-waveguide geometry to enhance field confinement and increase the chance that an activated defect overlaps strongly with the guided mode, but the broader takeaway is that monolithic processing provides a direct, fabrication-efficient route to on-chip routing without any heterogeneous transfer step \cite{li2021integration}. Experimentally, waveguide coupling is supported by collection maps showing signal at both the SQE site and the two out-couplers (Fig. \ref{fig: waveguide integration}j), along with matching spectra collected at the SQE and at the couplers (Fig. \ref{fig: waveguide integration}k--l) \cite{li2021integration}. A central quantitative message is that monolithic integration can, in principle, substantially improve coupling compared with stacked or hybrid schemes. Li \textit{et al.} provide an FDTD simulation benchmark that gives $\beta \approx 0.40$ for an ideal dipole at the center of the monolithic waveguide, compared with $\beta \approx 0.11$ for a surface case \cite{li2021integration}. However, Fig. \ref{fig: waveguide integration}i--l also make clear why this does not immediately translate into deterministic scalability. Because defects are generated stochastically, the defect's spatial offset from the field maximum and its dipole orientation relative to the waveguide axis are generally uncontrolled. In the representative device, polarization analysis indicates a dipole about $70^\circ$ relative to the waveguide axis, for which the authors note a simulated maximum $\beta \approx 0.37$, yet the experimentally estimated coupling is $\beta \approx 0.032$ (reported as a lower bound) \cite{li2021integration}.

Recent work across the field has therefore focused on reducing this random-placement bottleneck while retaining the practical strengths of the monolithic paradigm. One route is top-down, spatially programmable activation within hBN photonic circuits, for example, by using localized electron irradiation to create SQEs at chosen locations after (or alongside) hBN nanophotonic fabrication, enabling an elementary monolithic quantum photonic circuit with a deliberately placed SQE \cite{gerard2023top}. A complementary line of work pursues hybrid integration on mature dielectric PIC platforms such as SiN, where deterministic alignment procedures can be used to register an hBN SQE to a pre-defined guided mode and, in some cases, to align the SQE dipole relative to the waveguide axis \cite{elshaari2021deterministic, yamashita2025spatially}. A different approach is to eliminate transfer steps altogether by directly growing hBN films on SiN waveguide substrates, leveraging wafer-scale materials growth as an entry point to integrated hBN photonics \cite{glushkov2021direct}. Efficient coupling of hBN SPEs into guided optical modes has also been demonstrated using non-planar geometries such as tapered optical fibers, reaching reported system collection efficiencies on the order of $\sim 10\%$. This underscores that the main limitation in planar waveguides is typically not the existence of a guided channel, but rather deterministic SQE placement and dipole alignment relative to that channel \cite{schell2017coupling}.

Taken together, Fig. \ref{fig: waveguide integration}i--l, and these supporting demonstrations reinforce a central message for hBN waveguide integration. Monolithic hBN nanophotonics offers an elegant and widely adopted route to on-chip routing with a favorable best-case coupling landscape, but achieving high yield and uniform device performance will require deterministic control over defect creation, deterministic placement at the guided-mode field maximum, and dipole orientation.

\subsection{On-Chip Integration Methods}

\subsubsection{Transition Metal Dichalcogenides}
Bringing 2D-material quantum emitters into on-chip resonators requires solving two practical problems at once: placing the SQE where the cavity field is strongest, and ensuring the SQE can be tuned into spectral resonance without compromising single-photon performance. A clear example of how these challenges can be addressed in a scalable device geometry is given by Ma \textit{et al.} \cite{ma2022chip} 

In this work, a dry-transferred WSe$_2$ monolayer is integrated onto a Si$_3$N$_4$ microring so that strain-localized SQEs can interact with the resonator's evanescent whispering-gallery field (Fig. \ref{fig: on-chip photonics}a, b). The integration stack and geometry are deliberately engineered to make spatial alignment largely built in. Etched support regions and the ring sidewall help stabilize the monolayer while simultaneously generating localized strain sites that activate SQEs directly at the ring perimeter, precisely where the cavity mode intensity has an evanescent field that can be coupled to (Fig. \ref{fig: on-chip photonics}a). Once integrated, the ring's quasi-TE resonance spectrum provides a set of discrete optical states that the SQE can be matched to (Fig. \ref{fig: on-chip photonics}c), while patterned inner-sidewall scattering features provide a practical outcoupling pathway that converts circulating in-plane cavity power into collectable free-space radiation (Fig. \ref{fig: on-chip photonics}d). The authors validate genuine resonant light-matter interaction not only through spectral signatures but via a dynamical metric. When the SQE is tuned onto a cavity mode, the measured lifetime shortens relative to the off-resonant case, consistent with a Purcell-enhanced radiative rate (Fig. \ref{fig: on-chip photonics}e). Polarization-resolved spectroscopy shows that the cavity channel can also mediate chiral or polarization-selective emission tied to the microring's spin-orbit-locked evanescent field structure (Fig. \ref{fig: on-chip photonics}f), emphasizing that resonator integration can enable both rate enhancement and control of emitted photonic degrees of freedom.

A complementary on-chip resonator approach, useful to contrast with traveling-wave microrings, is to employ localized photonic-crystal cavity modes that can be engineered for strong field confinement, directional outcoupling, and tolerance-aware SQE placement. In this spirit, Noori \textit{et al.} introduce a cavity-integration concept specifically aimed at boosting the usable collected signal from TMD SQEs by combining a mechanically compliant 2D layer with a nanophotonic cavity whose field maximum is spatially well localized \cite{noori2016photonic}. Their analysis highlights two practical takeaways that map directly onto integration method considerations. First, vertical SQE position and nanometer-scale overlap with the cavity antinode can dominate performance, making the mechanics of the transferred monolayer, including conformality, sagging, spacing layers, and residue, as important as the optical design itself. Second, cavity-enabled extraction can be engineered to be highly directional, so that enhancement is not only a matter of increasing the radiative rate but also of redistributing emission into a collection-friendly channel. That is, whereas ring resonators naturally excel at providing a circuit-compatible guided mode with straightforward waveguide interfacing, localized photonic crystal-style approaches emphasize mode-volume-driven enhancement and engineered far-field extraction, at the cost of tighter sensitivity to placement, fabrication tolerances, and emitter-cavity detuning.

Taken together, Fig. \ref{fig: on-chip photonics}a-f present a complete resonator-integration story for TMD SQEs. Transfer-based placement and edge-defined activation deliver spatial overlap, engineered resonances and outcouplers make the cavity channel experimentally accessible, and lifetime and polarization measurements verify genuine cavity coupling. Robust optical enhancement still requires the SQE to stay spectrally aligned with the cavity mode, so tuning strategies and spectral stability become decisive for practical operation.

\subsubsection{Hexagonal Boron Nitride}

\begin{figure*}[t!] \centering
     \includegraphics[scale=0.82]{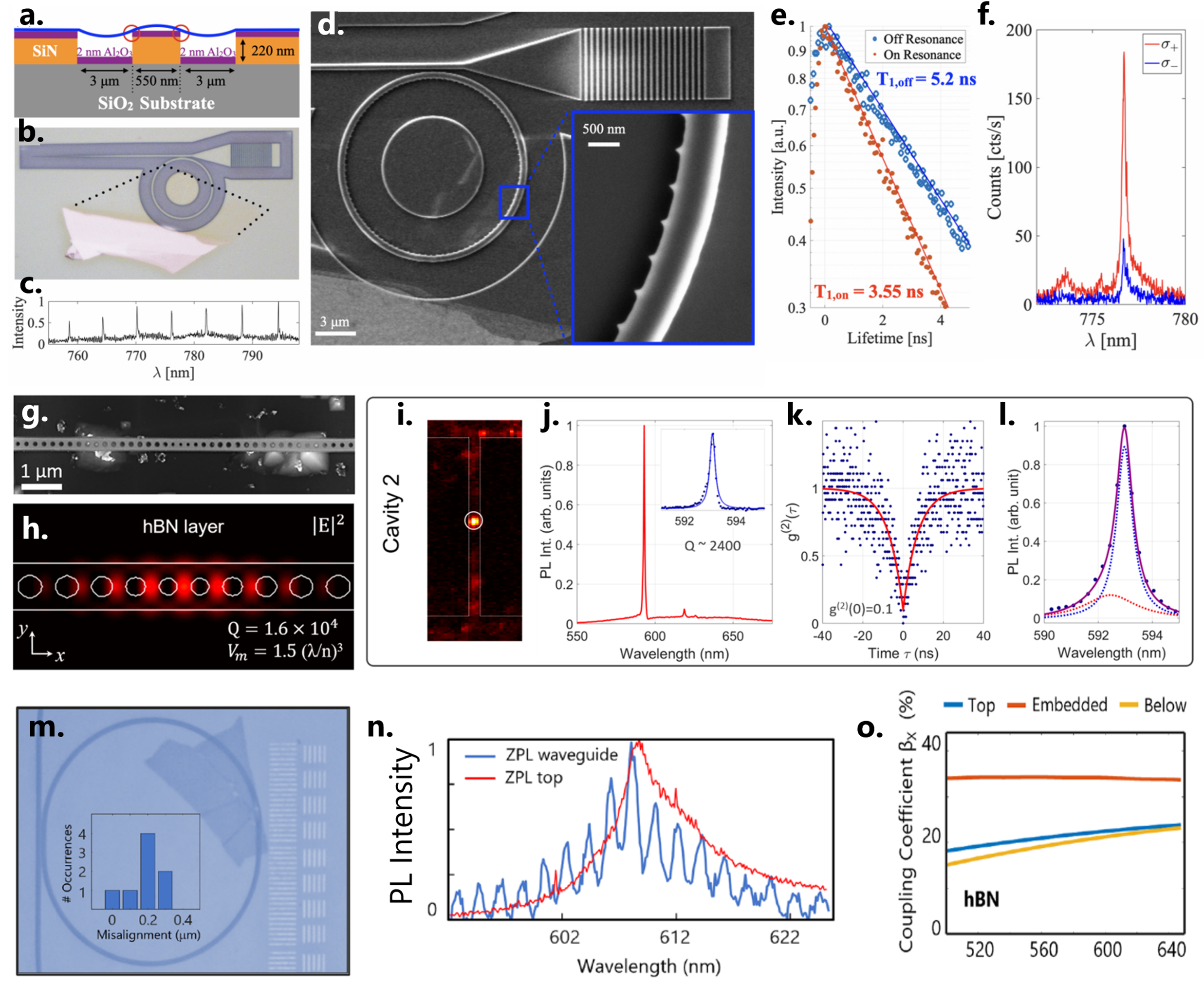}  
     \vspace{-0.1cm}
          \caption{\footnotesize \textbf{Photonic resonator integration for on-chip applications.} \textbf{(a)} Schematic cross-section of the hybrid WSe$_2$-Si$_3$N$_4$ microring platform showing the SiN ring geometry, air-trench release regions, and thin Al$_2$O$_3$ spacer used to support the transferred monolayer and promote evanescent coupling to the ring mode (adapted from \cite{ma2022chip} under the CC BY 4.0 license). \textbf{(b)} Optical micrograph of a fabricated microring with integrated waveguide/grating coupler and a dry-transferred WSe$_2$ flake covering most of the ring circumference to enable strain-localized SQE formation at the ring perimeter (adapted from \cite{ma2022chip} under the CC BY 4.0 license). \textbf{(c)} Representative quasi-TE microring transmission spectrum showing multiple whispering-gallery resonances across the SQE band with FSR and linewidth defining the cavity spectral structure used for tuning and enhancement (adapted from \cite{ma2022chip} under the CC BY 4.0 license). \textbf{(d)} SEM of the microring resonator highlighting the coupling region and engineered inner-sidewall scatterers. Inset shows a close-up of the periodic scattering features used to outcouple circulating cavity light to free space (adapted from \cite{ma2022chip} under the CC BY 4.0 license). \textbf{(e)} Time-resolved photoluminescence demonstrating resonant enhancement. SQE lifetime measured off-resonance is compared to on-resonance when tuned (via magnetic field) into a cavity mode, evidencing an increased radiative decay rate (adapted from \cite{ma2022chip} under the CC BY 4.0 license). \textbf{(f)} Polarization-resolved emission spectrum ($\sigma^+,\sigma^-$) from a cavity-coupled SQE, illustrating cavity-mediated readout and the polarization selectivity associated with the microring's evanescent field and scattering outcoupling (adapted from \cite{ma2022chip} under the CC BY 4.0 license).\textbf{(g)} SEM micrograph of a fabricated Si$_3$N$_4$ photonic crystal nanobeam cavity, showing the 1D hole lattice and tapered defect region that localizes the cavity mode (adapted from \cite{froch2020coupling}, Copyright 2020, American Chemical Society). \textbf{(h)} Simulated cavity field intensity $|E|^2$ for the nanobeam cavity with an hBN layer on top, illustrating strong field confinement at the cavity center and the expected spatial overlap between the cavity antinode and SQEs in the hBN film (adapted from \cite{froch2020coupling}, Copyright 2020, American Chemical Society). \textbf{(i)} Confocal photoluminescence map of a representative device ("cavity 2") showing a bright quantum emitter positioned near the cavity defect region where the optical field is maximal (adapted from \cite{froch2020coupling}, Copyright 2020, American Chemical Society). \textbf{(j)} Photoluminescence spectrum acquired at the coupled SQE location, showing spectral overlap between the SQE ZPL and a cavity resonance, with an inset highlighting the cavity resonance (adapted from \cite{froch2020coupling}, Copyright 2020, American Chemical Society). \textbf{(k)} Second order intensity autocorrelation of the cavity coupled emission, demonstrating antibunching and confirming single photon emission after integration (adapted from \cite{froch2020coupling}, Copyright 2020, American Chemical Society). \textbf{(l)} Spectral lineshape analysis of the coupled emission, where a multi-component fit separates SQE and cavity contributions to quantify cavity mediated enhancement and linewidth characteristics (adapted from \cite{froch2020coupling}, Copyright 2020, American Chemical Society). \textbf{(m)} Optical micrograph of a deterministically integrated hBN emitter-SiN microring resonator device, with an inset histogram summarizing the lateral placement (misalignment) accuracy achieved across multiple integrations (adapted from \cite{parto2022cavity} under the CC BY 4.0 license).\textbf{(n)} Photoluminescence spectra collected from the waveguide output and from free-space collection above the device, showing resonator-imprinted spectral features in the guided channel that evidence on-chip routing of SQE emission through the microring resonator (adapted from \cite{parto2022cavity} under the CC BY 4.0 license). \textbf{(o)} Simulated coupling coefficient $\beta$ versus wavelength for hBN SQEs positioned on top of, embedded within, or below the SiN photonic structure, illustrating how vertical placement in the stack governs coupling strength into the on-chip mode (adapted from \cite{parto2022cavity} under the CC BY 4.0 license).}
          \label{fig: on-chip photonics}
\end{figure*}

Hexagonal boron nitride offers a particularly flexible route to on-chip resonant enhancement because emitters can be hosted in thin, transferable flakes while the photonic circuitry is defined lithographically in a separate layer. In practice, the dominant integration challenge is not simply bringing the materials into contact, but ensuring that the SQE is positioned within a cavity field antinode and that the cross-sectional stack provides strong optical overlap without introducing excess loss or spectral instability. Two complementary demonstrations, one emphasizing statistically assembled cavity coupling and the other emphasizing deterministic device registration, illustrate how these requirements shape scalable integration workflows \cite{froch2020coupling, parto2022cavity}. 

Fr{\"o}ch \textit{et al.} couple hBN SQEs to lithographically defined Si$_3$N$_4$ one-dimensional photonic crystal nanobeam cavities, leveraging the small mode volume and well-localized field profile of the defect mode to obtain resonant enhancement \cite{froch2020coupling}. The platform itself is naturally compatible with array-scale fabrication because the nanobeam cavity geometry is compact and repeatable across a chip (Fig. \ref{fig: on-chip photonics}g). The key integration step is then establishing reliable emitter-mode overlap in the plane of the device. In this approach, spatial alignment is achieved statistically using a high density of SQEs in CVD-grown hBN and a cavity design whose antinode is spatially well defined, followed by confocal mapping to identify SQEs that lie close to the cavity center (Fig. \ref{fig: on-chip photonics}i). Optical simulations motivate why this geometry is effective by showing that, with the hBN layer placed on top of the nanobeam, the cavity field maximum remains concentrated at the defect while maintaining substantial overlap with the hBN film, with reported design values of $Q \sim 1.6\times10^4$, $V_m \sim 1.5(\lambda/n)^3$, and field overlap approaching $\sim40\%$ for thin hBN (Fig. \ref{fig: on-chip photonics}h) \cite{froch2020coupling}. Spectral coupling is established when a narrow hBN ZPL overlaps the cavity resonance, producing a cavity-shaped lineshape and loaded $Q$ values on the order of a few thousand in representative devices (Fig. \ref{fig: on-chip photonics}j) \cite{froch2020coupling}. Importantly for any resonant-enhancement strategy, the integrated emission retains single-photon character with strong antibunching (Fig. \ref{fig: on-chip photonics}k), and spectral decomposition of the coupled emission indicates photoluminescence enhancement factors reaching up to $\sim9\times$ in favorable cases (Fig. \ref{fig: on-chip photonics}l) \cite{froch2020coupling}. Together, these results highlight a highly scalable cavity platform in which the main reproducibility lever is yield, namely the probability of finding an SQE within the spatial capture region of the cavity antinode and with sufficient spectral proximity to a resonance.

A complementary integration philosophy is to remove the stochastic element of emitter-mode alignment by first locating SQEs and then fabricating the photonic structure around them. Parto \textit{et al.} demonstrate this deterministic workflow for hBN SQEs integrated with SiN microresonators and waveguides, using alignment markers to localize an SQE in a thin hBN flake and then patterning the SiN photonic layer so that the resonator mode is placed at the known SQE position (Fig. \ref{fig: on-chip photonics}m) \cite{parto2022cavity}. The misalignment statistics summarized in the figure inset emphasize the scalability of this approach because repeated devices can be fabricated with sub-micron placement precision, enabling arrays of emitter-resonator systems without relying on post-fabrication discovery. A particularly valuable consequence of this deterministic integration is that it directly supports circuit-level functionality. By comparing emission collected from above the device to emission collected from the waveguide output, the waveguide spectrum shows clear resonator-imprinted structure, demonstrating that emission is funneled through the microring resonator and routed on chip rather than being observed only in free-space collection (Fig. \ref{fig: on-chip photonics}n) \cite{parto2022cavity}. This distinction is central for integrated quantum photonics because it ties resonant enhancement to a guided, chip-compatible mode that can be interferometrically processed downstream.

Across both cavity paradigms, the cross-sectional placement of the hBN layer relative to the guided mode emerges as a dominant design variable, setting the achievable $\beta$ factor and the balance between optical overlap and SQE robustness. Parto \textit{et al.} explicitly quantify this trade space by comparing simulated coupling for SQEs located on top of, embedded within, or below the SiN photonic structure (Fig. \ref{fig: on-chip photonics}o) \cite{parto2022cavity}. Embedding generally maximizes spatial overlap and can yield the highest coupling into the guided mode, while placing hBN below the photonic layer can reduce interaction with etched sidewalls and processing residues, potentially preserving SQE stability at the cost of reduced overlap. In practice, these results motivate a set of integration considerations that apply broadly to hBN on-chip resonators: control of vertical spacing layers, minimization of scattering loss introduced by the transferred flake and interfaces, management of dipole orientation relative to the cavity polarization, and ensuring that any deterministic placement accuracy is proportionate with the lateral extent of the cavity antinode. When these constraints are addressed, hBN integration can move beyond proof-of-principle enhancement toward reproducible, waveguide-routed single-photon sources that are compatible with large-scale nanophotonic circuitry (Fig. \ref{fig: on-chip photonics}g-o) \cite{froch2020coupling, parto2022cavity}. 

Beyond these two representative demonstrations (Fig. \ref{fig: on-chip photonics}g--o), the broader hBN cavity-coupling literature reinforces three recurring integration lessons: scalable fabrication of high-Q resonators, methods for spectral matching in the presence of ZPL variability and spectral diffusion, and workflows that shift spatial alignment from a statistical search problem toward deterministic placement.

On the monolithic side, Kim \textit{et al.} establish that suspended photonic crystal cavities can be engineered directly from hBN with Q factors exceeding 2000, and importantly introduce an iterative, deterministic tuning knob using direct-write electron-beam induced etching that shifts the cavity resonance without strongly degrading Q \cite{kim2018photonic}. This framing is significant for on-chip resonator integration because it flips the usual tuning burden of rather than forcing a spectrally wandering SQE to match a fixed cavity, the cavity itself can be post-fabrication trimmed toward the emitter, providing a practical route to reproducible resonance alignment in arrays.

Building directly on that platform, Fr{\"o}ch \textit{et al.} demonstrate controlled Purcell enhancement for a cavity-coupled hBN SQE in a monolithic 1D photonic crystal cavity by tuning the SQE into resonance via cryogenic gas condensation. In resonance they report a strong ZPL intensity enhancement alongside a measurable lifetime reduction and infer a large Purcell factor, explicitly using combined intensity and lifetime metrics to quantify cavity-modified emission dynamics \cite{froch2022purcell}. In the context of integration methods, this work highlights that reproducible enhancement is not only a fabrication problem but also a tuning and measurement protocol problem, where on/off resonance comparisons provide the most unambiguous evidence of cavity-enhanced radiative emission.

From a process and scalability standpoint, Nonahal \textit{et al.} demonstrate that a wide variety of hBN nanophotonic components including tapered waveguides, microdisks, and 1D and 2D photonic crystal cavities can be fabricated from single-crystal hBN, with measured Q factors exceeding 4000 for 1D photonic crystal cavities in suspended architectures \cite{nonahal2023engineering}. A key takeaway for on-chip resonator integration is that once the photonic platform itself is monolithic in the SQE host, spatial overlap is intrinsically guaranteed, shifting the remaining bottleneck toward site-specific defect creation and spectral control.

Cranwell Schaeper \textit{et al.} address a different but equally practical integration bottleneck: yield and reproducibility of nanofabrication for large-area vdW photonic devices. Their double-etch masking and pattern-transfer protocol is designed to avoid common limitations of resist-based and metal-mask approaches, and they demonstrate functional hBN waveguides, ring resonators, and photonic crystal cavities across relevant spectral bands, including regimes associated with $V_B^-$ spin defects and visible quantum emitters \cite{schaeper2024double}. For resonator-coupled SQE devices, this style of fabrication advance matters because it reduces device-to-device variability in sidewall quality and optical loss, and it enables the kind of multi-device statistics that heterogeneous integration studies often rely on when SQEs are stochastically found rather than deterministically placed.

Wang \textit{et al.} provide a complementary device-design perspective by analyzing hBN microdisk whispering-gallery resonators as a route to relax spectral matching constraints. Because microdisk resonances can be engineered by choosing radial and azimuthal mode indices, the geometry offers a flexible way to access cavity modes across a broad ZPL distribution, which is attractive for hBN where ZPLs can span a wide wavelength range \cite{wang2021cavity}. While largely theoretical, the central takeaway for integration methodology is that cavity choice is itself a tuning strategy where resonator families with dense mode spectra and easily scalable geometry parameters can reduce the burden on active tuning, especially when paired with on-chip routing and filtering.

Taken together, these additional studies broaden the integration picture around Fig. \ref{fig: on-chip photonics}g--o. Heterogeneous approaches such as Fr{\"o}ch \textit{et al.} \cite{froch2020coupling} and Parto \textit{et al.} \cite{parto2022cavity} emphasize engineered overlap through stack design and deterministic placement, while monolithic hBN nanophotonics \cite{kim2018photonic, nonahal2023engineering, froch2022purcell, schaeper2024double} increasingly treat resonators as part of the SQE material system itself, enabling intrinsic spatial overlap and placing emphasis on reproducible fabrication and controllable spectral tuning. Across both tracks, the cross-sectional architecture remains a dominant design lever where maximizing $\beta$ into a guided mode requires pushing the SQE toward the field antinode and increasing optical confinement, but it must be balanced against fabrication processing-induced degradation, background auto-fluorescence, and charge-noise-driven spectral diffusion that can wash out resonant enhancement unless spectral matching can be maintained.

\subsection{Off-Chip Integration Methods}

\begin{figure*}[t!] \centering
     \includegraphics[scale=0.6]{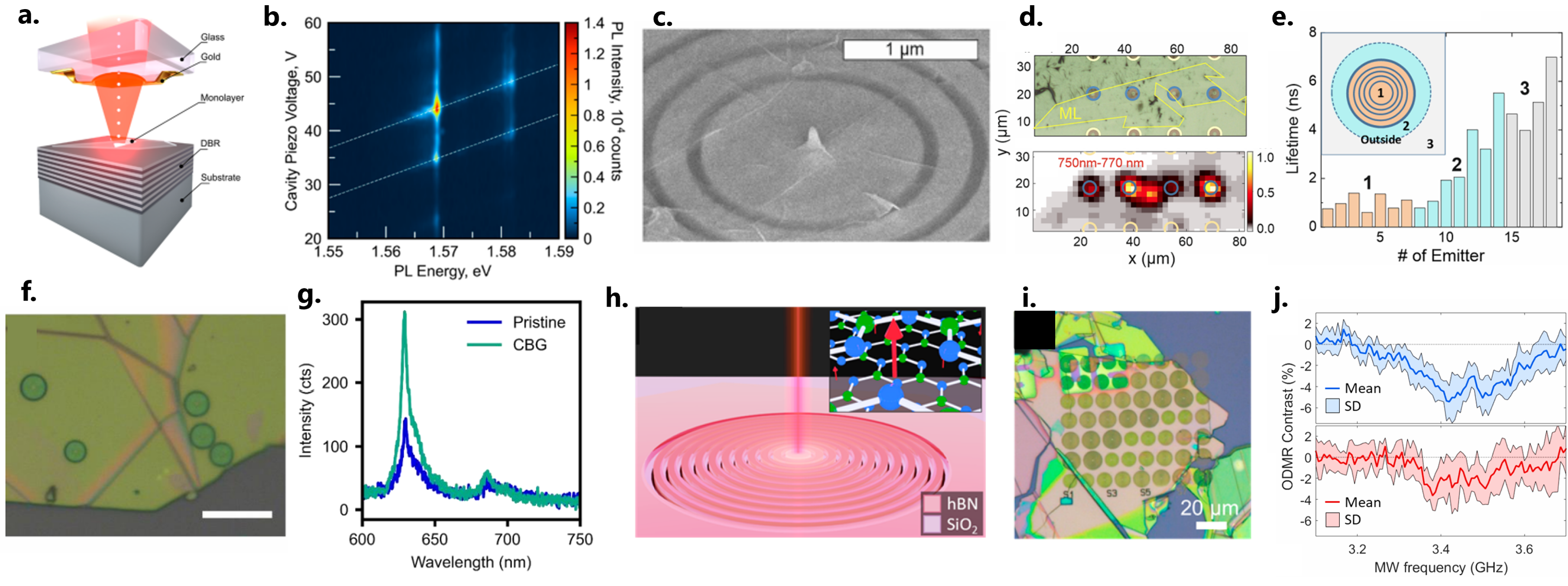}  
     \vspace{-0.5cm}
          \caption{\footnotesize \textbf{Photonic resonator integration for off-chip applications} \textbf{(a)} Open Fabry--Perot microcavity integration of a TMD SQE. Asymmetric open cavity with a WSe$_2$ monolayer on the bottom DBR mirror and a concave top mirror, enabling in situ positioning and cavity-length tuning for resonant enhancement and free-space outcoupling (adapted from \cite{drawer2023monolayer} under the CC BY 4.0 license). \textbf{(b)} Resonant enhancement map under cavity-length tuning. PL intensity (color) versus emission energy and cavity piezo voltage, showing cavity-mode branches crossing a narrow emitter line and producing a strong on-resonance brightness increase (adapted from \cite{drawer2023monolayer} under the CC BY 4.0 license). \textbf{(c)} Circular Bragg gratting (CBG) cavity with deterministic strain-site for SQE generation. SEM of a circular Bragg grating CBG cavity featuring a central nanopillar that strains an overlaid WSe$_2$ monolayer to localize an SQE near the cavity field maximum (adapted from \cite{iff2021purcell}, Copyright 2021, American Chemical Society). \textbf{(d)} Spatially correlated cavity-enhanced emission at device centers. Top: optical image indicating CBG cavity locations under the monolayer. Bottom: integrated PL map (750--770 nm window) showing emission hotspots co-located with the CBG centers, consistent with robust spatial overlap between SQE sites and the cavity mode (adapted from \cite{iff2021purcell}, Copyright 2021, American Chemical Society). \textbf{(e)} Lifetime statistics indicating Purcell enhancement in the cavity mode region. Measured lifetimes for many SQEs grouped by radial position relative to the CBG cavity (inset: regions 1--3), showing systematically shortened lifetimes for SQEs closest to the cavity center compared with outer regions (adapted from \cite{iff2021purcell}, Copyright 2021, American Chemical Society). \textbf{(f)} Deterministic circular-Bragg-grating (CBG) integration around pre-localized hBN SQEs. Optical micrograph of an exfoliated hBN flake showing multiple patterned CBG devices fabricated at the mapped positions of pre-screened SQEs using marker-referenced confocal mappings (adapted from \cite{liddle2026deterministic} under the CC BY 4.0 license). \textbf{(g)} Spectral evidence of cavity-enhanced emission after CBG fabrication. Representative PL spectra of the same hBN SQE before patterning (pristine) and after integration into a CBG cavity, showing increased ZPL intensity consistent with successful emitter-cavity mode overlap (adapted from \cite{liddle2026deterministic} under the CC BY 4.0 license). \textbf{(h)} Schematic of a CBG cavity patterned monolithically in hBN to funnel defect emission into a directed free-space mode, illustrated for $V_B^-$ spin defects (adapted from \cite{froch2021coupling}, Copyright 2021, American Chemical Society). \textbf{(i)} Optical micrograph of array-level implementation of an hBN flake patterned with a CBG cavity array (adapted from \cite{froch2021coupling}, Copyright 2021, American Chemical Society). \textbf{(j)} Cavity-enhanced optically-detected-magnetic-resonance (ODMR) readout enabled by improved collection. Comparison of ODMR performance (inside vs outside the CBG region), showing stronger optical readout (higher ODMR contrast) for SQEs coupled to the CBG cavity mode (adapted from \cite{froch2021coupling}, Copyright 2021, American Chemical Society).}
          \label{fig: off-chip photonics}
\end{figure*}

\subsubsection{Transition Metal Dichalcogenides}

Off-chip resonators provide an attractive integration route for 2D quantum emitters because they enhance the light-matter interaction while directing the emitted photons vertically out of the chip into highly directional, fiber-compatible free-space modes. This geometry can deliver strong Purcell enhancement while largely avoiding invasive nanofabrication of the SQE host, thereby shifting the main challenge from fabricating the cavity to achieving efficient spatial and spectral mode matching between the emitter and the resonator mode.

Figure \ref{fig: off-chip photonics}a, b highlight a representative open-access Fabry-P\'erot strategy that preserves the pristine 2D material while enabling in situ alignment. Drawer \textit{et al.} implement a plano-convex open microcavity where a WSe$_2$ monolayer is positioned on a high-reflectivity DBR bottom mirror and paired with a concave, Au-coated top mirror (Fig. \ref{fig: off-chip photonics}a) to define a well-confined free-space cavity mode \cite{drawer2023monolayer}. A key integration advantage of this architecture is that alignment is performed in situ, where nanopositioners enable lateral overlap of the cavity waist with a selected SQE site and tuning of the mirror separation so that a cavity resonance is brought onto the SQE ZPL \cite{drawer2023monolayer}. In practice, the approach benefits from hBN encapsulation and from using strain-localized SQE sites that can be identified optically (e.g., near flake wrinkles/edges) and then aligned to the cavity mode via cavity scanning \cite{drawer2023monolayer}. The resulting output is a directed, near-Gaussian mode explicitly motivated for efficient free-space collection and potential fiber coupling, at the expense of a mechanically more complex and alignment-sensitive platform relative to monolithic nanophotonics.

The corresponding evidence for resonant coupling is shown in Fig.\ref{fig: off-chip photonics}b, where the PL spectrum is mapped as the cavity length is scanned \cite{drawer2023monolayer}. As discrete cavity modes sweep across the SQE energy, the PL brightens sharply when a mode crosses the ZPL. On resonance, an intensity increase exceeding an order of magnitude is reported. In the off-chip context, Fig. \ref{fig: off-chip photonics}b emphasizes two practical benefits: (i) continuous spectral alignment that can accommodate SQE-to-SQE variability, and (ii) mode-selective, directed emission into the cavity output channel rather than broad free-space leaky modes. 

Beyond the specific example in Fig. \ref{fig: off-chip photonics}a-b, closely related open-cavity demonstrations help clarify the performance knobs and operating regimes of this approach. Flatten \textit{et al.} use an open-access plano-concave microcavity coupled to a WSe$_2$ SQE and emphasize that cavity-length scans and transverse-mode structure provide a practical diagnostic for optimizing lateral overlap between the SQE and the cavity waist where they report Purcell enhancement alongside a substantial increase in detected photon flux under cavity coupling \cite{flatten2018microcavity}. Separately, Drawer \textit{et al.} extend the same tune-the-cavity-through-the-emitter philosophy of Fig. \ref{fig: off-chip photonics}b to deterministically created WS$_2$ micro-dome SQEs integrated into an open Fabry-P\'erot cavity, demonstrating large brightness enhancement while also highlighting that phonon sidebands can contribute to off-resonant SQE-cavity coupling: an important nuance when interpreting cavity-induced enhancement in open geometries \cite{drawer2025tunable}.

In contrast to open Fabry-P\'erot systems that rely on in situ scanning for spatial matching, Fig. \ref{fig: off-chip photonics}c-e emphasize a circular Bragg grating (CBG) strategy that can enforce directed vertical radiation and can be engineered to address spatial alignment structurally. In Fig. \ref{fig: off-chip photonics}c, Iff \textit{et al.} employ a circular Bragg grating cavity with a central nanopillar such that a transferred WSe$_2$ monolayer forms a locally strained region near the cavity field maximum, enabling straightforward spatial alignment of a localized SQE to the resonant mode \cite{iff2021purcell}. Figure \ref{fig: off-chip photonics}d then supports the scalability of this workflow through hyperspectral mapping where emission hotspots repeatedly coincide with CBG centers when integrating over wavelength windows that overlap the cavity response, consistent with position-locked emitter-mode overlap across multiple devices rather than one-off alignment schemes \cite{iff2021purcell}. Figure \ref{fig: off-chip photonics}e provides quantitative evidence of radiative-rate modification by compiling lifetime statistics versus radial position, revealing systematically shortened lifetimes closest to the cavity center, this is consistent with Purcell enhancement when spatial overlap and spectral proximity are simultaneously achieved \cite{iff2021purcell}.

Two additional studies further contextualize Fig. \ref{fig: off-chip photonics}c-e as part of a broader free-space resonator-engineering toolbox for TMD SQEs. Duong \textit{et al.} demonstrate that circular Bragg grating structures can strongly reshape and enhance WSe$_2$ emission with pronounced directionality, reinforcing the CBG motif as a powerful route to vertical extraction \cite{duong2018enhanced}. Complementing these experimental demonstrations, Hekmati \textit{et al.} provide a design-centric perspective on dielectric CBG cavities for surface quantum emitters, a geometry naturally compatible with 2D materials, and outline how such structures can be optimized to simultaneously deliver high collection efficiency and Purcell enhancement into a well-defined free-space mode \cite{hekmati2023bullseye}.

\subsubsection{Hexagonal Boron Nitride}

In hBN, off-chip resonator integration has converged on a few complementary alignment strategies that trade mechanical tunability for deterministic fabrication, but all aim to convert intrinsically bright (yet spatially/spectrally variable) defect emission into a well-defined optical mode suitable for efficient free-space collection (Fig. \ref{fig: off-chip photonics}f-j). In the examples highlighted here, one route is to deterministically place a cavity around a pre-identified SQE (Fig. \ref{fig: off-chip photonics}f, g), while another is to use planar Bragg geometries to engineer directionality and collection across scalable device arrays, including for spin-defect ensembles (Fig. \ref{fig: off-chip photonics}h-j). Recent related work extends these ideas with tunable open Fabry-P\'erot microcavities bringing ``tune the cavity to the emitter'' flexibility to hBN.

A fabrication-native approach is illustrated in Fig. \ref{fig: off-chip photonics}f, where Liddle-Wesolowski \textit{et al.} implement a workflow that begins with activating and pre-screening visible hBN single-photon emitters, then locking their positions into a lithographic coordinate system using etched alignment markers and distortion-corrected confocal mapping \cite{liddle2026deterministic}. Rather than searching for post-fabrication overlap, the cavity is written where the SQE already is. Multiple circular Bragg grating cavities are patterned directly into the hBN flake at the mapped SQE coordinates, so spatial alignment is enforced by design \cite{liddle2026deterministic}. This map first, fabricate second strategy is particularly attractive for hBN because it preserves the ability to choose only the best SQEs in terms of brightness, stability, spectral features before committing to nanofabrication.

With that deterministic placement in hand, Fig. \ref{fig: off-chip photonics}g provides a direct, same-emitter before and after comparison showing the integration is optically functional. The ZPL becomes brighter after CBG fabrication, consistent with resonator-assisted funneling of emission into the collected mode \cite{liddle2026deterministic}. In Liddle-Wesolowski \textit{et al.}, representative devices show roughly two-fold ZPL enhancement ($\sim$2.2$\times$) and a corresponding increase in saturated count rate (e.g., from $\sim$81~kcps to $\sim$188~kcps, $\sim$2.3$\times$), underscoring that the primary benefit can be realized even without active mechanical tuning---by combining pre-fabrication metrology with a cavity geometry chosen to overlap the measured ZPL \cite{liddle2026deterministic}.

A closely related push toward deterministic, monolithic circular Bragg grating integration has also been demonstrated by Spencer \textit{et al.}, who combine site-specific SQE creation (via electron-beam irradiation for the B center in hBN) with circular Bragg grating cavities designed to shape the far-field \cite{spencer2023monolithic}. In the context of Fig. \ref{fig: off-chip photonics}f, g, this line of work reinforces the broader message: hBN is unusually compatible with workflows where both the SQE location and the photonic structure are engineered in the same material system, enabling single-emitter devices with multi-fold intensity gains (reported up to $\sim$6$\times$ in collected intensity for a 436~nm SQE) \cite{spencer2023monolithic}.

A different set of design constraints appears when the target is not a narrowband ZPL, but instead a broad, ensemble-like defect spectrum and spin readout. Figure \ref{fig: off-chip photonics}h highlights this regime through the monolithic hBN circular Bragg grating cavity platform of Fr{\"o}ch \textit{et al.}, motivated for the negatively charged boron vacancy ($V_\mathrm{B}^-$) \cite{froch2021coupling}. Here the key enhancement mechanism is deliberately framed as directionality/extraction enhancement rather than purely lifetime reduction, that is, the cavity geometry can improve collection over a wide bandwidth and for a wide range of dipole orientations, while remaining planar and therefore naturally compatible with vdW hosts \cite{froch2021coupling}.

That emphasis on scalability is confirmed by Fig. \ref{fig: off-chip photonics}i, where an exfoliated hBN flake is patterned into arrays of circular Bragg grating cavities with multiple geometric scalings on a single chip, so that different devices span different resonance conditions across the defect's broad emission band \cite{froch2021coupling}. Defects are then generated by ion irradiation, improving the odds of spatial overlap between SQEs and the cavity fields \cite{froch2021coupling}. Across these arrays, the measured photoluminescence enhancement reaches multi-fold levels (up to $\sim$6.5$\times$) for the best-matched designs, while still remaining appreciable across nearby scalings, an important practical point for broadband spin-defect emission where perfect spectral matching is less meaningful than robust, repeatable collection gains \cite{froch2021coupling}.

\begin{figure*}[t!] \centering
     \includegraphics[scale=1.0]{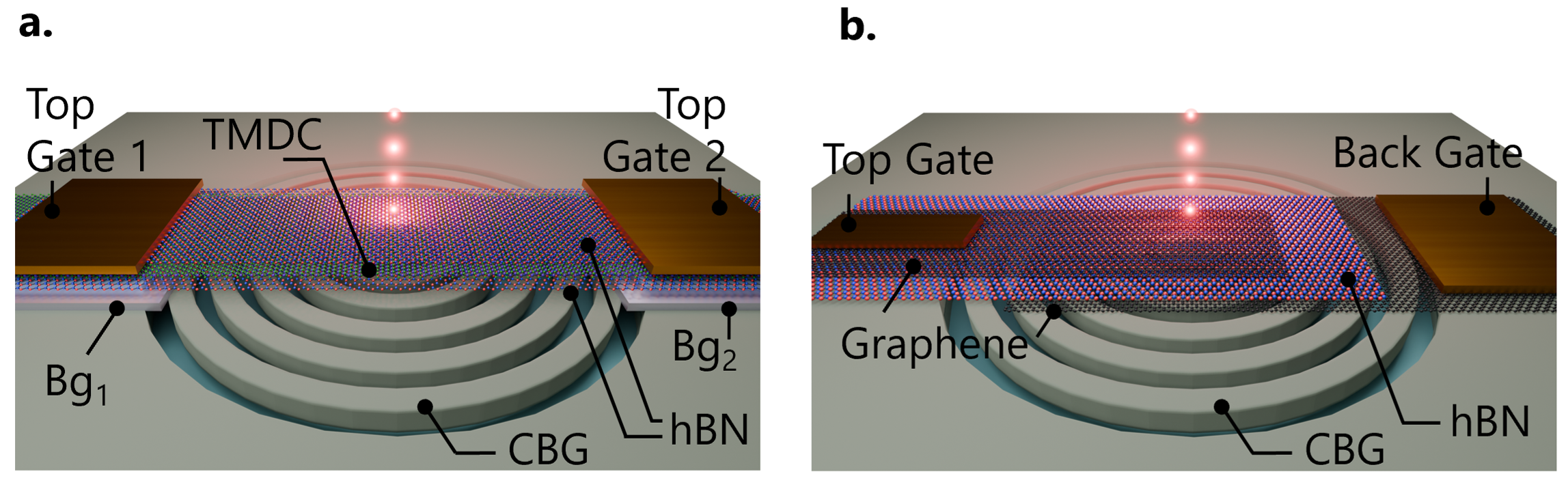}  
     \vspace{-0.5cm}
          \caption{\footnotesize \textbf{Co-designed electrical and photonic integration concepts for scalable 2D-material single-photon sources.} \textbf{(a)} Conceptual device stack for a localized SQE in an hBN-encapsulated TMD monolayer positioned over a circular Bragg grating cavity that funnels single-photon emission into a vertically directed, high-collection free-space mode. Split back gates (Bg$_1$, Bg$_2$) electrostatically tune and stabilize the local environment (e.g., charge control and Stark tuning), while patterned top gates/contacts enable local carrier injection and fast electrical modulation at the SQE site. \textbf{(b)} Analogous architecture for an hBN defect SQE integrated with a CBG cavity, where a graphene/hBN heterostructure provides a low-disorder electrostatic and contact platform where a top gate/contact and a back gate set the carrier density and field profile to prepare and stabilize the defect charge state, while the CBG converts the stabilized emission into a bright, directional, fiber-compatible output mode. Schematics are illustrative and not to scale.}
          \label{fig: outro}
\end{figure*}

Figure \ref{fig: off-chip photonics}j connects photonic integration to a functional spin metric by comparing ODMR readout inside versus outside the CBG region \cite{froch2021coupling}. By increasing the collected PL (and reducing the relative contribution of background), the cavity region yields improved ODMR performance, including higher contrast (reported around $\sim$5.4\% inside vs $\sim$3.6\% outside) and reduced scan-to-scan variability \cite{froch2021coupling}. This illustrates a particularly valuable off-chip advantage for hBN spin defects being that even when radiative-rate modification is difficult to resolve for broad ensembles, directional extraction alone can immediately translate into higher-fidelity optical spin readout.

Beyond the devices explicitly shown in Fig. \ref{fig: off-chip photonics}f-j, several works extend the same off-chip-resonator logic by using tunable open Fabry-P\'erot cavities, bringing a hBN analogue of spectral alignment by cavity-length tuning demonstrated for TMDs in Fig. \ref{fig: off-chip photonics}a, b. Vogl \textit{et al.} integrate hBN SQEs into a compact tunable microcavity and report cavity-induced spectral reshaping consistent with Purcell-assisted brightening and filtering, including excited-state lifetime shortening, linewidth narrowing, and improved single-photon purity through suppression of off-resonant background \cite{vogl2019compact}. H\"au{\ss}ler \textit{et al.} push this further with a fiber-based open cavity offering broad spatial and spectral tunability, reporting very large ZPL spectral enhancement (up to $\sim$50$\times$) and strong linewidth reduction attributed to cavity funneling, along with cavity-assisted photoluminescence-excitation spectroscopy that leverages simultaneous enhancement of excitation and collection channels \cite{haussler2021tunable}. More recently, Maier \textit{et al.} mitigate a practical bottleneck of open-cavity coupling, namely flake-induced height variations and associated scattering/round-trip loss, by isolating thin, membrane-like hBN structures that host single defects and inserting them into a fiber Fabry--Pérot cavity, yielding room-temperature spectral enhancement up to $\sim$100$\times$ \cite{maier2025extracting}. Collectively, these open-cavity studies reinforce a central message that in hBN, where ZPLs can be widely distributed across devices, tunability is often the most direct route to reliable spectral alignment and strong mode selectivity.

A complementary hybrid photonics direction is exemplified by work that brings a thin hBN film into intimate contact with dielectric whispering-gallery resonators (e.g., Si$_3$N$_4$ microdisks) via pick-and-place transfer \cite{proscia2020microcavity}. As the film conforms and wraps around the disk, strain can activate defects preferentially near the resonator perimeter, naturally placing SQEs within the whispering-gallery-mode volume and enabling cavity-enhanced collection without etching the hBN host \cite{proscia2020microcavity}. In the spirit of Fig. \ref{fig: off-chip photonics}f-j, this modular approach highlights a third pathway: rather than patterning cavities into hBN or mechanically tuning an external cavity, one can leverage mature dielectric microresonators to provide high-quality factor photonics while using hBN as the SQE layer, allowing the SQE host and the photonic system to be optimized semi-independently.

\section*{Outlook and Conclusion}

A recurring theme throughout this review is that electrical addressability and photonic mode engineering solve different bottlenecks, and that the highest-performing 2D-material single-photon sources will ultimately require both, implemented in a co-designed device stack rather than as add-on improvements. On the electronic side, gates and contacts provide (i) a pathway to triggered operation via electrical injection and fast modulation, and (ii) electrostatic stabilization that suppresses charge-noise-induced spectral wandering and blinking, enabling narrower linewidths and more reproducible operation. On the photonic side, cavities and engineered radiation channels convert intrinsically isotropic emission into a single, collectable optical mode, simultaneously improving extraction and, when desired, increasing the radiative rate through Purcell enhancement. These complementary roles are already explicit in the conceptual presentation of literature and in the device progress surveyed across electrical injection, charge control, and cavity/waveguide platforms.

Figure \ref{fig: outro} is a natural synthesis of this logic into two architectures that could plausibly become workhorse building blocks for scalable 2D quantum photonics. In Fig. \ref{fig: outro}a, an hBN-encapsulated TMD SQE stack is paired with back gates that electrostatically dope and stabilize the local environment and top gates/contacts that deliver current to drive emission directly at the localized site, while a circular Bragg grating cavity funnels the optical output into a vertically directed, fiber-compatible free-space mode. In practice, this architecture is compelling because it enables a single device to co-optimize three requirements that are often pursued separately and can be mutually constraining: (i) on-demand operation via current injection and fast switching, (ii) spectral stabilization and tunability through charge-state control, depletion, and Stark tuning, and (iii) efficient optical interfacing by funneling optical emission into a well-defined, high-collection spatial mode. The outlook opportunity is that, once combined, these elements can be optimized jointly, for example, gate geometries that minimize local field noise can be chosen alongside cavity designs that maximize extraction without forcing invasive processing near the SQE.


Analogous to the TMD case, Fig.~\ref{fig: outro}b illustrates a graphene-hBN heterostructure that provides a clean, tunable electrostatic environment for hBN SQEs, where back gates set the local carrier density and field profile, while top graphene gates/contacts enable current injection or controlled charge-state preparation at the defect. When integrated with a CBG cavity, the electrically stabilized emission is efficiently outcoupled into a bright, directional mode, with gating ensuring repeatable spectral stability and the cavity providing enhanced extraction.

As the field pushes towards deploying sources, the bottleneck shifts toward system-level figures of merit that cannot be achieved by electronics or photonics alone. For example, a cavity can boost collection dramatically, but if the SQE's transition frequency jitters faster than the cavity linewidth, the usable enhancement and indistinguishability collapse. Conversely, electrical stabilization can narrow and stabilize the line, but without a controlled optical mode the collected fraction remains small and device-to-device packaging becomes ad hoc. The architectures in Fig.~\ref{fig: outro} are therefore best viewed as platform-level proposals for bringing 2D quantum emitters into the same engineering paradigm that has driven progress in other solid-state single-photon technologies.

In summary, this article surveyed progress in electronic and photonic integration of SQEs in 2D materials, emphasizing how device engineering is steadily transforming these SQEs from laboratory demonstrations into candidate building blocks for scalable quantum photonics. Electrically, the field has converged on a small set of robust vdW device archetypes, vertical tunneling junctions, double-barrier LEDs, and gate-defined lateral junctions, that enable carrier injection while preserving localized emission, and has increasingly leveraged gating not just for tuning but for charge stabilization that directly targets spectral diffusion and blinking. In parallel, Photonic integration has matured from proof-of-principle waveguide routing to resonator-enhanced emission in both on-chip and off-chip geometries, with CBG-style structures standing out as a practical pathway to highly directional, packaging-compatible free-space collection.

A key takeaway is that the next phase of progress is unlikely to be dominated by electronics or photonics alone, but by their intentional combination and co-design. The performance metrics that matter most for quantum technologies, high single-photon purity, high brightness into a single mode, spectral stability, and ultimately photon indistinguishability, are intrinsically coupled to both the local electrostatic environment and the optical mode structure. Figure \ref{fig: outro} captures this convergence by proposing integrated stacks where gates prepare and stabilize the SQE while a CBG cavity defines the emitted mode. Such co-designed platforms point toward ``turnkey'' 2D-material quantum-light engines with electrically addressable, photonics-ready SQE that can be triggered on demand, tuned and stabilized against noise, and efficiently interfaced with fibers and photonic circuits. Reaching that goal will require continued advances in deterministic SQE placement, low-noise gating and materials passivation, cavity-compatible stabilization protocols, and packaging-aware photonic design, but the trajectory summarized here suggests that simultaneous electronic and photonic integration is the most direct route to scalable, deployable SQE in the 2D-material platform.

\begin{acknowledgments} 
\noindent S.D.P., S.D., and G.M. gratefully acknowledge support from
NSF Award No. ECCS-2032272 and the NSF Quantum Foundry through Q-AMASE-i program Award No.
DMR-1906325. 
\end{acknowledgments}

\section*{Author Contributions}
\noindent S.D.P and S.D. contributed equally to this work. S.D., L.J., and S.D.P obtained copyright permissions. K.P. performed simulations of PN-based quantum LEDs. S.D.P and S.D. wrote the manuscript. G.M. supervised the project. 

\section*{Data Availability}
\noindent Data sharing is not applicable to this article as no new data were created or analyzed in this study.

\section*{Competing Interests}
\noindent The authors declare no conflicts of interest.


\bibliographystyle{unsrt}
\bibliography{Biblio}

\begin{thebibliography}{100}

\bibitem{fournier2021position}
Clarisse Fournier, Alexandre Plaud, S{\'e}bastien Roux, Aur{\'e}lie Pierret, Michael Rosticher, Kenji Watanabe, Takashi Taniguchi, St{\'e}phanie Buil, Xavier Qu{\'e}lin, Julien Barjon, et~al.
\newblock Position-controlled quantum emitters with reproducible emission wavelength in hexagonal boron nitride.
\newblock {\em Nature communications}, 12(1):3779, 2021.

\bibitem{ahmed2025nanoindentation}
Safa~L Ahmed, Lukas Harsch, Csongor Imre, Ioannis Karapatzakis, Luis Kussi, Jeremias Resch, Marcel Schrodin, Ines H{\"a}usler, Tolga Wagner, Christoph~T Koch, et~al.
\newblock Nanoindentation for tailored single-photon emitters in hbn: Influence of annealing on defect stability.
\newblock {\em ACS nano}, 19(41):36302--36312, 2025.

\bibitem{doan2025near}
Sean Doan, Sahil~D Patel, Yilin Chen, Jordan A~Gusdorff Turiansky, E~Mark, Luis Villagomez, Luka Jevremovic, Nicholas Lewis, Kenji Watanabe, Takashi Taniguchi, et~al.
\newblock Near-infrared quantum emission from oxygen-related defects in hbn.
\newblock {\em arXiv preprint arXiv:2512.16197}, 2025.

\bibitem{he2015single}
Yu-Ming He, Genevieve Clark, John~R. Schaibley, Yu~He, Ming-Cheng Chen, Yu-Jia Wei, Xing Ding, Qiang Zhang, Wang Yao, Xiaodong Xu, Chao-Yang Lu, and Jian-Wei Pan.
\newblock Single quantum emitters in monolayer semiconductors.
\newblock {\em Nature Nanotechnology}, 10(6):497--502, 2015.

\bibitem{srivastava2015optically}
Ajit Srivastava, Meinrad Sidler, Adrien~V. Allain, Dominik~S. Lembke, Andras Kis, and Atac Imamoglu.
\newblock Optically active quantum dots in monolayer wse2.
\newblock {\em Nature Nanotechnology}, 10(6):491--496, 2015.

\bibitem{koperski2015single}
M.~Koperski, K.~Nogajewski, A.~Arora, V.~Cherkez, P.~Mallet, J.-Y. Veuillen, J.~Marcus, P.~Kossacki, and M.~Potemski.
\newblock Single photon emitters in exfoliated wse2 structures.
\newblock {\em Nature Nanotechnology}, 10(6):503--506, 2015.

\bibitem{tonndorf2015single}
Philipp Tonndorf, Robert Schmidt, Robert Schneider, Johannes Kern, Michele Buscema, Gary~A. Steele, Andres Castellanos-Gomez, Herre S.~J. van~der Zant, Steffen Michaelis~de Vasconcellos, and Rudolf Bratschitsch.
\newblock Single-photon emission from localized excitons in an atomically thin semiconductor.
\newblock {\em Optica}, 2(4):347--352, 2015.

\bibitem{branny2017deterministic}
Artur Branny, Santosh Kumar, Rapha{\"e}l Proux, and Brian~D Gerardot.
\newblock Deterministic strain-induced arrays of quantum emitters in a two-dimensional semiconductor.
\newblock {\em Nature communications}, 8(1):15053, 2017.

\bibitem{palaciosberraquero2017large}
Carmen Palacios-Berraquero, Dhiren~M. Kara, Alejandro R.-P. Montblanch, Matteo Barbone, Pawel Latawiec, Duhee Yoon, Anna~K. Ott, Marko Loncar, Andrea~C. Ferrari, and Mete Atature.
\newblock Large-scale quantum-emitter arrays in atomically thin semiconductors.
\newblock {\em Nature Communications}, 8:15093, 2017.

\bibitem{kern2016nanoscale}
Johannes Kern, Iris Niehues, Philipp Tonndorf, Robert Schmidt, Daniel Wigger, Robert Schneider, Torsten Stiehm, Steffen Michaelis~de Vasconcellos, Doris~E. Reiter, Tilmann Kuhn, and Rudolf Bratschitsch.
\newblock Nanoscale positioning of single-photon emitters in atomically thin wse2.
\newblock {\em Advanced Materials}, 28(33):7101--7105, 2016.

\bibitem{parto2021defect}
Kamyar Parto, Shaimaa~I Azzam, Kaustav Banerjee, and Galan Moody.
\newblock Defect and strain engineering of monolayer wse2 enables site-controlled single-photon emission up to 150 k.
\newblock {\em Nature communications}, 12(1):3585, 2021.

\bibitem{shepard2017nanobubble}
Gabriella~D. Shepard, Obafunso~A. Ajayi, Xiangzhi Li, X.-Y. Zhu, James Hone, and Stefan Strauf.
\newblock Nanobubble induced formation of quantum emitters in monolayer semiconductors.
\newblock {\em 2D Materials}, 4(2):021019, 2017.

\bibitem{darlington2020imaging}
Thomas~P. Darlington, Christian Carmesin, Matthias Florian, Emanuil Yanev, Obafunso Ajayi, Jenny Ardelean, Daniel~A. Rhodes, Augusto Ghiotto, Andrey Krayev, Kenji Watanabe, Takashi Taniguchi, Jeffrey~W. Kysar, Abhay~N. Pasupathy, James~C. Hone, Frank Jahnke, Nicholas~J. Borys, and P.~James Schuck.
\newblock Imaging strain-localized excitons in nanoscale bubbles of monolayer wse2 at room temperature.
\newblock {\em Nature Nanotechnology}, 15:854--860, 2020.

\bibitem{so2021electrically}
Jae-Pil So, Ha-Reem Kim, Hyeonjun Baek, Kwang-Yong Jeong, Hoo-Cheol Lee, Woong Huh, Yoon~Seok Kim, Kenji Watanabe, Takashi Taniguchi, Jungkil Kim, et~al.
\newblock Electrically driven strain-induced deterministic single-photon emitters in a van der waals heterostructure.
\newblock {\em Science Advances}, 7(43):eabj3176, 2021.

\bibitem{clark2016single}
Genevieve Clark, John~R Schaibley, Jason Ross, Takashi Taniguchi, Kenji Watanabe, Joshua~R Hendrickson, Shin Mou, Wang Yao, and Xiaodong Xu.
\newblock Single defect light-emitting diode in a van der waals heterostructure.
\newblock {\em Nano letters}, 16(6):3944--3948, 2016.

\bibitem{lenferink2022tunable}
Erik~J Lenferink, Trevor LaMountain, Teodor~K Stanev, Ethan Garvey, Kenji Watanabe, Takashi Taniguchi, and Nathaniel~P Stern.
\newblock Tunable emission from localized excitons deterministically positioned in monolayer p--n junctions.
\newblock {\em ACS Photonics}, 9(9):3067--3074, 2022.

\bibitem{grzeszczyk2024electroluminescence}
Magdalena Grzeszczyk, Kristina Vaklinova, Kenji Watanabe, Takashi Taniguchi, Konstantin~S Novoselov, and Maciej Koperski.
\newblock Electroluminescence from pure resonant states in hbn-based vertical tunneling junctions.
\newblock {\em Light: Science \& Applications}, 13(1):155, 2024.

\bibitem{aharonovich2016solid}
Igor Aharonovich, Dirk Englund, and Milos Toth.
\newblock Solid-state single-photon emitters.
\newblock {\em Nature photonics}, 10(10):631--641, 2016.

\bibitem{o2009photonic}
Jeremy~L O'brien, Akira Furusawa, and Jelena Vu{\v{c}}kovi{\'c}.
\newblock Photonic quantum technologies.
\newblock {\em Nature photonics}, 3(12):687--695, 2009.

\bibitem{aharonovich2022quantum}
Igor Aharonovich, Jean-Philippe Tetienne, and Milos Toth.
\newblock Quantum emitters in hexagonal boron nitride.
\newblock {\em Nano Letters}, 22(23):9227--9235, 2022.

\bibitem{kianinia2017robust}
Mehran Kianinia, Blake Regan, Sherif~Abdulkader Tawfik, Toan~Trong Tran, Michael~J Ford, Igor Aharonovich, and Milos Toth.
\newblock Robust solid-state quantum system operating at 800 k.
\newblock {\em Acs Photonics}, 4(4):768--773, 2017.

\bibitem{cassabois2016hexagonal}
Guillaume Cassabois, Pierre Valvin, and Bernard Gil.
\newblock Hexagonal boron nitride is an indirect bandgap semiconductor.
\newblock {\em Nature photonics}, 10(4):262--266, 2016.

\bibitem{caldwell2019photonics}
Joshua~D Caldwell, Igor Aharonovich, Guillaume Cassabois, James~H Edgar, Bernard Gil, and Dimitri~N Basov.
\newblock Photonics with hexagonal boron nitride.
\newblock {\em Nature Reviews Materials}, 4(8):552--567, 2019.

\bibitem{gale2022site}
Angus Gale, Chi Li, Yongliang Chen, Kenji Watanabe, Takashi Taniguchi, Igor Aharonovich, and Milos Toth.
\newblock Site-specific fabrication of blue quantum emitters in hexagonal boron nitride.
\newblock {\em Acs Photonics}, 9(6):2170--2177, 2022.

\bibitem{gottscholl2020initialization}
Andreas Gottscholl, Mehran Kianinia, Victor Soltamov, Sergei Orlinskii, Georgy Mamin, Carlo Bradac, Christian Kasper, Klaus Krambrock, Andreas Sperlich, Milos Toth, et~al.
\newblock Initialization and read-out of intrinsic spin defects in a van der waals crystal at room temperature.
\newblock {\em Nature materials}, 19(5):540--545, 2020.

\bibitem{tran2016quantum}
Toan~Trong Tran, Kerem Bray, Michael~J Ford, Milos Toth, and Igor Aharonovich.
\newblock Quantum emission from hexagonal boron nitride monolayers.
\newblock {\em Nature nanotechnology}, 11(1):37--41, 2016.

\bibitem{tran2016robust}
Toan~Trong Tran, Christopher Elbadawi, Daniel Totonjian, Charlene~J Lobo, Gabriele Grosso, Hyowon Moon, Dirk~R Englund, Michael~J Ford, Igor Aharonovich, and Milos Toth.
\newblock Robust multicolor single photon emission from point defects in hexagonal boron nitride.
\newblock {\em ACS nano}, 10(8):7331--7338, 2016.

\bibitem{jungwirth2016temperature}
Nicholas~R Jungwirth, Brian Calderon, Yanxin Ji, Michael~G Spencer, Michael~E Flatt{\'e}, and Gregory~D Fuchs.
\newblock Temperature dependence of wavelength selectable zero-phonon emission from single defects in hexagonal boron nitride.
\newblock {\em Nano letters}, 16(10):6052--6057, 2016.

\bibitem{grosso2017tunable}
Gabriele Grosso, Hyowon Moon, Benjamin Lienhard, Sajid Ali, Dmitri~K Efetov, Marco~M Furchi, Pablo Jarillo-Herrero, Michael~J Ford, Igor Aharonovich, and Dirk Englund.
\newblock Tunable and high-purity room temperature single-photon emission from atomic defects in hexagonal boron nitride.
\newblock {\em Nature communications}, 8(1):1--8, 2017.

\bibitem{luo2018deterministic}
Yue Luo, Gabriella~D Shepard, Jenny~V Ardelean, Daniel~A Rhodes, Bumho Kim, Katayun Barmak, James~C Hone, and Stefan Strauf.
\newblock Deterministic coupling of site-controlled quantum emitters in monolayer wse2 to plasmonic nanocavities.
\newblock {\em Nature nanotechnology}, 13(12):1137--1142, 2018.

\bibitem{zhao2021site}
Huan Zhao, Michael~T Pettes, Yu~Zheng, and Han Htoon.
\newblock Site-controlled telecom-wavelength single-photon emitters in atomically-thin mote2.
\newblock {\em Nature communications}, 12(1):6753, 2021.

\bibitem{gan2022large}
Lin Gan, Danyang Zhang, Ruiling Zhang, Qiyao Zhang, Hao Sun, Yongzhuo Li, and Cun-Zheng Ning.
\newblock Large-scale, high-yield laser fabrication of bright and pure single-photon emitters at room temperature in hexagonal boron nitride.
\newblock {\em ACS nano}, 16(9):14254--14261, 2022.

\bibitem{li2021scalable}
Chi Li, Noah Mendelson, Ritika Ritika, YongLiang Chen, Zai-Quan Xu, Milos Toth, and Igor Aharonovich.
\newblock Scalable and deterministic fabrication of quantum emitter arrays from hexagonal boron nitride.
\newblock {\em Nano Letters}, 21(8):3626--3632, 2021.

\bibitem{xu2021creating}
Xiaohui Xu, Zachariah~O Martin, Demid Sychev, Alexei~S Lagutchev, Yong~P Chen, Takashi Taniguchi, Kenji Watanabe, Vladimir~M Shalaev, and Alexandra Boltasseva.
\newblock Creating quantum emitters in hexagonal boron nitride deterministically on chip-compatible substrates.
\newblock {\em Nano letters}, 21(19):8182--8189, 2021.

\bibitem{o2024transfer}
Dante~J O'Hara, Hsun-Jen Chuang, Kathleen~M McCreary, Mehmet~A Noyan, Sung-Joon Lee, Enrique~D Cobas, and Berend~T Jonker.
\newblock Transfer of hexagonal boron nitride quantum emitters onto arbitrary substrates with zero thermal budget.
\newblock {\em APL Materials}, 12(7), 2024.

\bibitem{froch2020coupling}
Johannes~E Fr{\"o}ch, Sejeong Kim, Noah Mendelson, Mehran Kianinia, Milos Toth, and Igor Aharonovich.
\newblock Coupling hexagonal boron nitride quantum emitters to photonic crystal cavities.
\newblock {\em ACS nano}, 14(6):7085--7091, 2020.

\bibitem{li2021integration}
Chi Li, Johannes~E Froch, Milad Nonahal, Thinh~N Tran, Milos Toth, Sejeong Kim, and Igor Aharonovich.
\newblock Integration of hbn quantum emitters in monolithically fabricated waveguides.
\newblock {\em ACS photonics}, 8(10):2966--2972, 2021.

\bibitem{parto2022cavity}
Kamyar Parto, Shaimaa~I Azzam, Nicholas Lewis, Sahil~D Patel, Sammy Umezawa, Kenji Watanabe, Takashi Taniguchi, and Galan Moody.
\newblock Cavity-enhanced 2d material quantum emitters deterministically integrated with silicon nitride microresonators.
\newblock {\em Nano Letters}, 22(23):9748--9756, 2022.

\bibitem{patel2024surface}
Sahil~D Patel, Kamyar Parto, Michael Choquer, Nicholas Lewis, Sammy Umezawa, Landon Hellman, Daniella Polishchuk, and Galan Moody.
\newblock Surface acoustic wave cavity optomechanics with atomically thin h-bn and wse 2 single-photon emitters.
\newblock {\em PRX Quantum}, 5(1):010330, 2024.

\bibitem{sakib2024purcell}
Mashnoon~Alam Sakib, Brandon Triplett, William Harris, Naveed Hussain, Alexander Senichev, Melika Momenzadeh, Joshua Bocanegra, Polina Vabishchevich, Ruqian Wu, Alexandra Boltasseva, et~al.
\newblock Purcell-induced bright single photon emitters in hexagonal boron nitride.
\newblock {\em Nano Letters}, 24(40):12390--12397, 2024.

\bibitem{yamashita2025deterministic}
Daiki Yamashita, Masaki Yumoto, Aiko Narazaki, and Makoto Okano.
\newblock Deterministic integration of hbn single-photon emitters on sin waveguides via femtosecond laser processing.
\newblock {\em arXiv preprint arXiv:2504.19477}, 2025.

\bibitem{kim2018photonic}
Sejeong Kim, Johannes~E Fr{\"o}ch, Joe Christian, Marcus Straw, James Bishop, Daniel Totonjian, Kenji Watanabe, Takashi Taniguchi, Milos Toth, and Igor Aharonovich.
\newblock Photonic crystal cavities from hexagonal boron nitride.
\newblock {\em Nature communications}, 9(1):2623, 2018.

\bibitem{stern2022room}
Hannah~L Stern, Qiushi Gu, John Jarman, Simone Eizagirre~Barker, Noah Mendelson, Dipankar Chugh, Sam Schott, Hoe~H Tan, Henning Sirringhaus, Igor Aharonovich, et~al.
\newblock Room-temperature optically detected magnetic resonance of single defects in hexagonal boron nitride.
\newblock {\em Nature communications}, 13(1):618, 2022.

\bibitem{patel2024room}
Raj~N Patel, Rebecca~EK Fishman, Tzu-Yung Huang, Jordan~A Gusdorff, David~A Fehr, David~A Hopper, S~Alex Breitweiser, Benjamin Porat, Michael~E Flatt{\'e}, and Lee~C Bassett.
\newblock Room temperature dynamics of an optically addressable single spin in hexagonal boron nitride.
\newblock {\em Nano letters}, 24(25):7623--7628, 2024.

\bibitem{chejanovsky2021single}
Nathan Chejanovsky, Amlan Mukherjee, Jianpei Geng, Yu-Chen Chen, Youngwook Kim, Andrej Denisenko, Amit Finkler, Takashi Taniguchi, Kenji Watanabe, Durga Bhaktavatsala~Rao Dasari, et~al.
\newblock Single-spin resonance in a van der waals embedded paramagnetic defect.
\newblock {\em Nature materials}, 20(8):1079--1084, 2021.

\bibitem{gao2022nuclear}
Xingyu Gao, Sumukh Vaidya, Kejun Li, Peng Ju, Boyang Jiang, Zhujing Xu, Andres E~Llacsahuanga Allcca, Kunhong Shen, Takashi Taniguchi, Kenji Watanabe, et~al.
\newblock Nuclear spin polarization and control in hexagonal boron nitride.
\newblock {\em Nature Materials}, 21(9):1024--1028, 2022.

\bibitem{gao2021high}
Xingyu Gao, Boyang Jiang, Andres~E Llacsahuanga~Allcca, Kunhong Shen, Mohammad~A Sadi, Abhishek~B Solanki, Peng Ju, Zhujing Xu, Pramey Upadhyaya, Yong~P Chen, et~al.
\newblock High-contrast plasmonic-enhanced shallow spin defects in hexagonal boron nitride for quantum sensing.
\newblock {\em Nano Letters}, 21(18):7708--7714, 2021.

\bibitem{stern2024quantum}
Hannah~L Stern, Carmem M.~Gilardoni, Qiushi Gu, Simone Eizagirre~Barker, Oliver~FJ Powell, Xiaoxi Deng, Stephanie~A Fraser, Louis Follet, Chi Li, Andrew~J Ramsay, et~al.
\newblock A quantum coherent spin in hexagonal boron nitride at ambient conditions.
\newblock {\em Nature Materials}, 23(10):1379--1385, 2024.

\bibitem{fournier2023two}
Clarisse Fournier, S{\'e}bastien Roux, Kenji Watanabe, Takashi Taniguchi, St{\'e}phanie Buil, Julien Barjon, Jean-Pierre Hermier, and Aymeric Delteil.
\newblock Two-photon interference from a quantum emitter in hexagonal boron nitride.
\newblock {\em Physical Review Applied}, 19(4):L041003, 2023.

\bibitem{gottscholl2021room}
Andreas Gottscholl, Matthias Diez, Victor Soltamov, Christian Kasper, Andreas Sperlich, Mehran Kianinia, Carlo Bradac, Igor Aharonovich, and Vladimir Dyakonov.
\newblock Room temperature coherent control of spin defects in hexagonal boron nitride.
\newblock {\em Science Advances}, 7(14):eabf3630, 2021.

\bibitem{baber2021excited}
Simon Baber, Ralph Nicholas~Edward Malein, Prince Khatri, Paul~Steven Keatley, Shi Guo, Freddie Withers, Andrew~J Ramsay, and Isaac~J Luxmoore.
\newblock Excited state spectroscopy of boron vacancy defects in hexagonal boron nitride using time-resolved optically detected magnetic resonance.
\newblock {\em Nano Letters}, 22(1):461--467, 2021.

\bibitem{alzahrani2024negatively}
Yahya~A Alzahrani and Masfer Alkahtani.
\newblock Negatively charged boron-vacancy defect in hexagonal boron nitride nanoparticles.
\newblock {\em Applied Physics Letters}, 124(17), 2024.

\bibitem{robertson2023detection}
Islay~O Robertson, Sam~C Scholten, Priya Singh, Alexander~J Healey, Fernando Meneses, Philipp Reineck, Hiroshi Abe, Takeshi Ohshima, Mehran Kianinia, Igor Aharonovich, et~al.
\newblock Detection of paramagnetic spins with an ultrathin van der waals quantum sensor.
\newblock {\em ACS nano}, 17(14):13408--13417, 2023.

\bibitem{gong2023coherent}
Ruotian Gong, Guanghui He, Xingyu Gao, Peng Ju, Zhongyuan Liu, Bingtian Ye, Erik~A Henriksen, Tongcang Li, and Chong Zu.
\newblock Coherent dynamics of strongly interacting electronic spin defects in hexagonal boron nitride.
\newblock {\em Nature Communications}, 14(1):3299, 2023.

\bibitem{palacios2018atomically}
Carmen Palacios-Berraquero.
\newblock Atomically-thin quantum light emitting diodes.
\newblock In {\em Quantum confined excitons in 2-dimensional materials}, pages 71--89. Springer, 2018.

\bibitem{schwarz2016electrically}
S~Schwarz, Aleksey Kozikov, Freddie Withers, JK~Maguire, AP~Foster, S~Dufferwiel, Lee Hague, MN~Makhonin, LR~Wilson, AK~Geim, et~al.
\newblock Electrically pumped single-defect light emitters in wse2.
\newblock {\em 2D Materials}, 3(2):025038, 2016.

\bibitem{park2024narrowband}
Gyuna Park, Ivan Zhigulin, Hoyoung Jung, Jake Horder, Karin Yamamura, Yerin Han, Hyunje Cho, Hyeon-Woo Jeong, Kenji Watanabe, Takashi Taniguchi, et~al.
\newblock Narrowband electroluminescence from color centers in hexagonal boron nitride.
\newblock {\em Nano Letters}, 24(48):15268--15274, 2024.

\bibitem{wang2025low}
Xiaoran Wang, Jialong Li, Ruxin Liu, Shuo Zhang, Yun Yao, Siyuan Wang, Xiangyi Wang, Yaning Liang, Xiao Wang, Xuechao Yu, et~al.
\newblock Low-background single-photon emission and multiwavelength electroluminescence from carbon-doped hexagonal boron nitride.
\newblock {\em Chemistry of Materials}, 37(22):9048--9056, 2025.

\bibitem{zhigulin2025electrical}
Ivan Zhigulin, Gyuna Park, Karin Yamamura, Kenji Watanabe, Takashi Taniguchi, Milos Toth, Jonghwan Kim, and Igor Aharonovich.
\newblock Electrical generation of color centers in hexagonal boron nitride.
\newblock {\em ACS Applied Materials \& Interfaces}, 17(16):24129--24136, 2025.

\bibitem{schuler2020electrically}
Bruno Schuler, Katherine~A Cochrane, Christoph Kastl, Edward~S Barnard, Edward Wong, Nicholas~J Borys, Adam~M Schwartzberg, D~Frank Ogletree, F~Javier~Garc{\'\i}a de~Abajo, and Alexander Weber-Bargioni.
\newblock Electrically driven photon emission from individual atomic defects in monolayer ws2.
\newblock {\em Science advances}, 6(38):eabb5988, 2020.

\bibitem{guo2023electrically}
Shi Guo, Savvas Germanis, Takashi Taniguchi, Kenji Watanabe, Freddie Withers, and Isaac~J Luxmoore.
\newblock Electrically driven site-controlled single photon source.
\newblock {\em ACS photonics}, 10(8):2549--2555, 2023.

\bibitem{howarth2024electroluminescent}
James Howarth, Kristina Vaklinova, Magdalena Grzeszczyk, Giulio Baldi, Lee Hague, Marek Potemski, Kostya~S Novoselov, Aleksey Kozikov, and Maciej Koperski.
\newblock Electroluminescent vertical tunneling junctions based on wse2 monolayer quantum emitter arrays: Exploring tunability with electric and magnetic fields.
\newblock {\em Proceedings of the National Academy of Sciences}, 121(23):e2401757121, 2024.

\bibitem{yu2024electrically}
Mihyang Yu, Jeonghan Lee, Kenji Watanabe, Takashi Taniguchi, and Jieun Lee.
\newblock Electrically pumped h-bn single-photon emission in van der waals heterostructure.
\newblock {\em ACS nano}, 19(1):504--511, 2024.

\bibitem{ng2007physics}
Kwok~Kwok Ng and Simon~M Sze.
\newblock {\em Physics of semiconductor devices}.
\newblock Wiley-Interscience Hoboken, NJ, 2007.

\bibitem{schulman2018contact}
Daniel~S Schulman, Andrew~J Arnold, and Saptarshi Das.
\newblock Contact engineering for 2d materials and devices.
\newblock {\em Chemical Society Reviews}, 47(9):3037--3058, 2018.

\bibitem{prakash2017understanding}
Abhijith Prakash, Hesameddin Ilatikhameneh, Peng Wu, and Joerg Appenzeller.
\newblock Understanding contact gating in schottky barrier transistors from 2d channels.
\newblock {\em Scientific reports}, 7(1):12596, 2017.

\bibitem{wang2015ultrafast}
Haining Wang, Changjian Zhang, and Farhan Rana.
\newblock Ultrafast dynamics of defect-assisted electron--hole recombination in monolayer mos2.
\newblock {\em Nano letters}, 15(1):339--345, 2015.

\bibitem{wu2016defects}
Zhangting Wu, Zhongzhong Luo, Yuting Shen, Weiwei Zhao, Wenhui Wang, Haiyan Nan, Xitao Guo, Litao Sun, Xinran Wang, Yumeng You, et~al.
\newblock Defects as a factor limiting carrier mobility in wse2: A spectroscopic investigation.
\newblock {\em Nano Research}, 9(12):3622--3631, 2016.

\bibitem{jung2019transferred}
Younghun Jung, Min~Sup Choi, Ankur Nipane, Abhinandan Borah, Bumho Kim, Amirali Zangiabadi, Takashi Taniguchi, Kenji Watanabe, Won~Jong Yoo, James Hone, et~al.
\newblock Transferred via contacts as a platform for ideal two-dimensional transistors.
\newblock {\em Nature Electronics}, 2(5):187--194, 2019.

\bibitem{english2016improved}
Chris~D English, Gautam Shine, Vincent~E Dorgan, Krishna~C Saraswat, and Eric Pop.
\newblock Improved contacts to mos2 transistors by ultra-high vacuum metal deposition.
\newblock {\em Nano letters}, 16(6):3824--3830, 2016.

\bibitem{wang2019van}
Yan Wang, Jong~Chan Kim, Ryan~J Wu, Jenny Martinez, Xiuju Song, Jieun Yang, Fang Zhao, Andre Mkhoyan, Hu~Young Jeong, and Manish Chhowalla.
\newblock Van der waals contacts between three-dimensional metals and two-dimensional semiconductors.
\newblock {\em Nature}, 568(7750):70--74, 2019.

\bibitem{shen2021ultralow}
Pin-Chun Shen, Cong Su, Yuxuan Lin, Ang-Sheng Chou, Chao-Ching Cheng, Ji-Hoon Park, Ming-Hui Chiu, Ang-Yu Lu, Hao-Ling Tang, Mohammad~Mahdi Tavakoli, et~al.
\newblock Ultralow contact resistance between semimetal and monolayer semiconductors.
\newblock {\em Nature}, 593(7858):211--217, 2021.

\bibitem{li2023approaching}
Weisheng Li, Xiaoshu Gong, Zhihao Yu, Liang Ma, Wenjie Sun, Si~Gao, {\c{C}}a{\u{g}}{\i}l K{\"o}ro{\u{g}}lu, Wenfeng Wang, Lei Liu, Taotao Li, et~al.
\newblock Approaching the quantum limit in two-dimensional semiconductor contacts.
\newblock {\em Nature}, 613(7943):274--279, 2023.

\bibitem{zhang2019two}
Xu~Zhang, Jes{\'u}s Grajal, Jose~Luis Vazquez-Roy, Ujwal Radhakrishna, Xiaoxue Wang, Winston Chern, Lin Zhou, Yuxuan Lin, Pin-Chun Shen, Xiang Ji, et~al.
\newblock Two-dimensional mos2-enabled flexible rectenna for wi-fi-band wireless energy harvesting.
\newblock {\em Nature}, 566(7744):368--372, 2019.

\bibitem{yang2020ultrafast}
Sung~Jin Yang, Kyu-Tae Park, Jaeho Im, Sungjae Hong, Yangjin Lee, Byung-Wook Min, Kwanpyo Kim, and Seongil Im.
\newblock Ultrafast 27 ghz cutoff frequency in vertical wse2 schottky diodes with extremely low contact resistance.
\newblock {\em Nature communications}, 11(1):1574, 2020.

\bibitem{akbari2022lifetime}
Hamidreza Akbari, Souvik Biswas, Pankaj~Kumar Jha, Joeson Wong, Benjamin Vest, and Harry~A Atwater.
\newblock Lifetime-limited and tunable quantum light emission in h-bn via electric field modulation.
\newblock {\em Nano Letters}, 22(19):7798--7803, 2022.

\bibitem{white2022electrical}
Simon~JU White, Tieshan Yang, Nikolai Dontschuk, Chi Li, Zai-Quan Xu, Mehran Kianinia, Alastair Stacey, Milos Toth, and Igor Aharonovich.
\newblock Electrical control of quantum emitters in a van der waals heterostructure.
\newblock {\em Light: Science \& Applications}, 11(1):186, 2022.

\bibitem{chakraborty2019electrical}
Chitraleema Chakraborty, Nicholas~R Jungwirth, Gregory~D Fuchs, and A~Nick Vamivakas.
\newblock Electrical manipulation of the fine-structure splitting of wse 2 quantum emitters.
\newblock {\em Physical Review B}, 99(4):045308, 2019.

\bibitem{baek2020highly}
Hyeonjun Baek, Mauro Brotons-Gisbert, Zhe~Xian Koong, Aidan Campbell, Markus Rambach, Kenji Watanabe, Takashi Taniguchi, and Brian~D Gerardot.
\newblock Highly energy-tunable quantum light from moir{\'e}-trapped excitons.
\newblock {\em Science advances}, 6(37):eaba8526, 2020.

\bibitem{noh2018stark}
Gichang Noh, Daebok Choi, Jin-Hun Kim, Dong-Gil Im, Yoon-Ho Kim, Hosung Seo, and Jieun Lee.
\newblock Stark tuning of single-photon emitters in hexagonal boron nitride.
\newblock {\em Nano letters}, 18(8):4710--4715, 2018.

\bibitem{scavuzzo2019electrically}
Alessio Scavuzzo, Shai Mangel, Ji-Hoon Park, Sanghyup Lee, Dinh Loc~Duong, Christian Strelow, Alf Mews, Marko Burghard, and Klaus Kern.
\newblock Electrically tunable quantum emitters in an ultrathin graphene--hexagonal boron nitride van der waals heterostructure.
\newblock {\em Applied Physics Letters}, 114(6), 2019.

\bibitem{nikolay2019very}
Niko Nikolay, Noah Mendelson, Nikola Sadzak, Florian B{\"o}hm, Toan~Trong Tran, Bernd Sontheimer, Igor Aharonovich, and Oliver Benson.
\newblock Very large and reversible stark-shift tuning of single emitters in layered hexagonal boron nitride.
\newblock {\em Physical Review Applied}, 11(4):041001, 2019.

\bibitem{mendelson2019engineering}
Noah Mendelson, Zai-Quan Xu, Toan~Trong Tran, Mehran Kianinia, John Scott, Carlo Bradac, Igor Aharonovich, and Milos Toth.
\newblock Engineering and tuning of quantum emitters in few-layer hexagonal boron nitride.
\newblock {\em ACS nano}, 13(3):3132--3140, 2019.

\bibitem{xia2019room}
Yang Xia, Quanwei Li, Jeongmin Kim, Wei Bao, Cheng Gong, Sui Yang, Yuan Wang, and Xiang Zhang.
\newblock Room-temperature giant stark effect of single photon emitter in van der waals material.
\newblock {\em Nano letters}, 19(10):7100--7105, 2019.

\bibitem{zhigulin2023stark}
Ivan Zhigulin, Jake Horder, Viktor Iv{\'a}dy, Simon~JU White, Angus Gale, Chi Li, Charlene~J Lobo, Milos Toth, Igor Aharonovich, and Mehran Kianinia.
\newblock Stark effect of blue quantum emitters in hexagonal boron nitride.
\newblock {\em Physical Review Applied}, 19(4):044011, 2023.

\bibitem{paralikis2025tunable}
Athanasios Paralikis, Pawe{\l} Wyborski, Pietro Metuh, Niels Gregersen, and Battulga Munkhbat.
\newblock Tunable and low-noise wse 2 quantum emitters for quantum photonics.
\newblock {\em PRX Quantum}, 6(4):040339, 2025.

\bibitem{hotger2021gate}
Alexander Hotger, Julian Klein, Katja Barthelmi, Lukas Sigl, Florian Sigger, Wolfgang Manner, Samuel Gyger, Matthias Florian, Michael Lorke, Frank Jahnke, et~al.
\newblock Gate-switchable arrays of quantum light emitters in contacted monolayer mos2 van der waals heterodevices.
\newblock {\em Nano Letters}, 21(2):1040--1046, 2021.

\bibitem{chen2023gate}
Yuan Chen, Haidong Liang, Leyi Loh, Yiwei Ho, Ivan Verzhbitskiy, Kenji Watanabe, Takashi Taniguchi, Michel Bosman, Andrew~A Bettiol, and Goki Eda.
\newblock Gate-tunable bound exciton manifolds in monolayer mose2.
\newblock {\em Nano Letters}, 23(10):4456--4463, 2023.

\bibitem{yu2022electrical}
Mihyang Yu, Donggyu Yim, Hosung Seo, and Jieun Lee.
\newblock Electrical charge control of h-bn single photon sources.
\newblock {\em 2D Materials}, 9(3):035020, 2022.

\bibitem{steiner2025current}
Corinne Steiner, Rebecca Rahmel, Frank Volmer, Rika Windisch, Lars~H Janssen, Patricia Pesch, Kenji Watanabe, Takashi Taniguchi, Florian Libisch, Bernd Beschoten, et~al.
\newblock Current-induced brightening of vacancy-related emitters in hexagonal boron nitride.
\newblock {\em Physical Review Research}, 7(3):L032037, 2025.

\bibitem{fraunie2025charge}
Jules Frauni{\'e}, Tristan Clua-Provost, S{\'e}bastien Roux, Zhao Mu, Adrien Delpoux, Gr{\'e}gory Seine, Delphine Lagarde, Kenji Watanabe, Takashi Taniguchi, Xavier Marie, et~al.
\newblock Charge state tuning of spin defects in hexagonal boron nitride.
\newblock {\em Nano Letters}, 25(14):5836--5842, 2025.

\bibitem{chakraborty2017quantum}
Chitraleema Chakraborty, Kenneth~M Goodfellow, Sajal Dhara, Anthony Yoshimura, Vincent Meunier, and A~Nick Vamivakas.
\newblock Quantum-confined stark effect of individual defects in a van der waals heterostructure.
\newblock {\em Nano Letters}, 17(4):2253--2258, 2017.

\bibitem{mukherjee2020electric}
Arunabh Mukherjee, Chitraleema Chakraborty, Liangyu Qiu, and A~Nick Vamivakas.
\newblock Electric field tuning of strain-induced quantum emitters in wse2.
\newblock {\em AIP Advances}, 10(7), 2020.

\bibitem{cai2024charge}
Hongbing Cai, Abdullah Rasmita, Ruihua He, Zhaowei Zhang, Qinghai Tan, Disheng Chen, Naizhou Wang, Zhao Mu, John~JH Eng, Yongzhi She, et~al.
\newblock Charge-depletion-enhanced wse2 quantum emitters on gold nanogap arrays with near-unity quantum efficiency.
\newblock {\em Nature Photonics}, 18(8):842--847, 2024.

\bibitem{wyborski2025toward}
Pawe{\l} Wyborski, Athanasios Paralikis, Pietro Metuh, Martin~A Jacobsen, Christian Ruiz, Niels Gregersen, and Battulga Munkhbat.
\newblock Toward triggered generation of indistinguishable single-photons from mote $ \_2 $ quantum emitters.
\newblock {\em arXiv preprint arXiv:2508.20743}, 2025.

\bibitem{wu2025modulation}
Frances Camille~M Wu, Shang-Hsuan Wu, Bin Fang, Xintong Li, Jadon Zheng, Jean Anne~C Incorvia, and Edward~T Yu.
\newblock Modulation of single photon emission from suspended 1l wse2 under electrostatically induced strain.
\newblock {\em Nano Letters}, 2025.

\bibitem{kim2022high}
Gwangwoo Kim, Hyong~Min Kim, Pawan Kumar, Mahfujur Rahaman, Christopher~E Stevens, Jonghyuk Jeon, Kiyoung Jo, Kwan-Ho Kim, Nicholas Trainor, Haoyue Zhu, et~al.
\newblock High-density, localized quantum emitters in strained 2d semiconductors.
\newblock {\em ACS nano}, 16(6):9651--9659, 2022.

\bibitem{chakraborty2015voltage}
Chitraleema Chakraborty, Laura Kinnischtzke, Kenneth~M Goodfellow, Ryan Beams, and A~Nick Vamivakas.
\newblock Voltage-controlled quantum light from an atomically thin semiconductor.
\newblock {\em Nature nanotechnology}, 10(6):507--511, 2015.

\bibitem{peyskens2019integration}
Fr{\'e}d{\'e}ric Peyskens, Chitraleema Chakraborty, Muhammad Muneeb, Dries Van~Thourhout, and Dirk Englund.
\newblock Integration of single photon emitters in 2d layered materials with a silicon nitride photonic chip.
\newblock {\em Nature communications}, 10(1):4435, 2019.

\bibitem{errando2021resonance}
Carlos Errando-Herranz, Eva Sch{\"o}ll, Rapha{\"e}l Picard, Micaela Laini, Samuel Gyger, Ali~W Elshaari, Art Branny, Ulrika Wennberg, Sebastien Barbat, Thibaut Renaud, et~al.
\newblock Resonance fluorescence from waveguide-coupled, strain-localized, two-dimensional quantum emitters.
\newblock {\em ACS photonics}, 8(4):1069--1076, 2021.

\bibitem{white2019atomically}
Daniel White, Artur Branny, Robert~J Chapman, Rapha{\"e}l Picard, Mauro Brotons-Gisbert, Andreas Boes, Alberto Peruzzo, Cristian Bonato, and Brian~D Gerardot.
\newblock Atomically-thin quantum dots integrated with lithium niobate photonic chips.
\newblock {\em Optical Materials Express}, 9(2):441--448, 2019.

\bibitem{tonndorf2017chip}
Philipp Tonndorf, Osvaldo Del Pozo-Zamudio, Nico Gruhler, Johannes Kern, Robert Schmidt, Alexander~I Dmitriev, Anatoly~P Bakhtinov, Alexander~I Tartakovskii, Wolfram Pernice, Steffen Michaelis~de Vasconcellos, et~al.
\newblock On-chip waveguide coupling of a layered semiconductor single-photon source.
\newblock {\em Nano letters}, 17(9):5446--5451, 2017.

\bibitem{gerard2023top}
Domitille G{\'e}rard, Michael Rosticher, Kenji Watanabe, Takashi Taniguchi, Julien Barjon, St{\'e}phanie Buil, Jean-Pierre Hermier, and Aymeric Delteil.
\newblock Top-down integration of an hbn quantum emitter in a monolithic photonic waveguide.
\newblock {\em Applied Physics Letters}, 122(26), 2023.

\bibitem{elshaari2021deterministic}
Ali~W Elshaari, Anas Skalli, Samuel Gyger, Martin Nurizzo, Lucas Schweickert, Iman Esmaeil~Zadeh, Mikael Svedendahl, Stephan Steinhauer, and Val Zwiller.
\newblock Deterministic integration of hbn emitter in silicon nitride photonic waveguide.
\newblock {\em Advanced Quantum Technologies}, 4(6):2100032, 2021.

\bibitem{yamashita2025spatially}
Daiki Yamashita, Masaki Yumoto, Aiko Narazaki, and Makoto Okano.
\newblock Spatially deterministic integration of hbn single-photon emitters on sin waveguides via femtosecond laser processing.
\newblock {\em Advanced Optical Materials}, 13(27):e01231, 2025.

\bibitem{glushkov2021direct}
Evgenii Glushkov, Noah Mendelson, Andrey Chernev, Ritika Ritika, Martina Lihter, Reza~R Zamani, Jean Comtet, Vytautas Navikas, Igor Aharonovich, and Aleksandra Radenovic.
\newblock Direct growth of hexagonal boron nitride on photonic chips for high-throughput characterization.
\newblock {\em ACS Photonics}, 8(7):2033--2040, 2021.

\bibitem{schell2017coupling}
Andreas~W Schell, Hideaki Takashima, Toan~Trong Tran, Igor Aharonovich, and Shigeki Takeuchi.
\newblock Coupling quantum emitters in 2d materials with tapered fibers.
\newblock {\em Acs Photonics}, 4(4):761--767, 2017.

\bibitem{ma2022chip}
Yichen Ma, Haoqi Zhao, Na~Liu, Zihe Gao, Seyed~Sepehr Mohajerani, Licheng Xiao, James Hone, Liang Feng, and Stefan Strauf.
\newblock On-chip spin-orbit locking of quantum emitters in 2d materials for chiral emission.
\newblock {\em Optica}, 9(8):953--958, 2022.

\bibitem{noori2016photonic}
Yasir~J Noori, Yameng Cao, Jonathan Roberts, Christopher Woodhead, Ramon Bernardo-Gavito, Peter Tovee, and Robert~J Young.
\newblock Photonic crystals for enhanced light extraction from 2d materials.
\newblock {\em Acs Photonics}, 3(12):2515--2520, 2016.

\bibitem{froch2022purcell}
Johannes~E Fr{\"o}ch, Chi Li, Yongliang Chen, Milos Toth, Mehran Kianinia, Sejeong Kim, and Igor Aharonovich.
\newblock Purcell enhancement of a cavity-coupled emitter in hexagonal boron nitride.
\newblock {\em Small}, 18(2):2104805, 2022.

\bibitem{nonahal2023engineering}
Milad Nonahal, Chi Li, Haoran Ren, Lesley Spencer, Mehran Kianinia, Milos Toth, and Igor Aharonovich.
\newblock Engineering quantum nanophotonic components from hexagonal boron nitride.
\newblock {\em Laser \& Photonics Reviews}, 17(8):2300019, 2023.

\bibitem{schaeper2024double}
Otto~Cranwell Schaeper, Lesley Spencer, Dominic Scognamiglio, Waleed El-Sayed, Benjamin Whitefield, Jake Horder, Nathan Coste, Paul Barclay, Milos Toth, Anastasiia Zalogina, et~al.
\newblock Double etch method for the fabrication of nanophotonic devices from van der waals materials.
\newblock {\em ACS Photonics}, 11(12):5446--5452, 2024.

\bibitem{wang2021cavity}
Yanan Wang, Jaesung Lee, Jesse Berezovsky, and Philip X-L Feng.
\newblock Cavity quantum electrodynamics design with single photon emitters in hexagonal boron nitride.
\newblock {\em Applied Physics Letters}, 118(24), 2021.

\bibitem{drawer2023monolayer}
Jens-Christian Drawer, Victor~Nikolaevich Mitryakhin, Hangyong Shan, Sven Stephan, Moritz Gittinger, Lukas Lackner, Bo~Han, Gilbert Leibeling, Falk Eilenberger, Rounak Banerjee, et~al.
\newblock Monolayer-based single-photon source in a liquid-helium-free open cavity featuring 65\% brightness and quantum coherence.
\newblock {\em Nano letters}, 23(18):8683--8689, 2023.

\bibitem{iff2021purcell}
Oliver Iff, Quirin Buchinger, Magdalena Mocza{\l}a-Dusanowska, Martin Kamp, Simon Betzold, Marcelo Davanco, Kartik Srinivasan, Sefaattin Tongay, Carlos Ant{\'o}n-Solanas, Sven Hofling, et~al.
\newblock Purcell-enhanced single photon source based on a deterministically placed wse2 monolayer quantum dot in a circular bragg grating cavity.
\newblock {\em Nano letters}, 21(11):4715--4720, 2021.

\bibitem{liddle2026deterministic}
James Liddle-Wesolowski, Otto~Cranwell Schaeper, Nathan Coste, Benjamin Whitefield, Evan Williams, Helen Zhi~Jie Zeng, Mehran Kianinia, Anastasiia Zalogina, and Igor Aharonovich.
\newblock Deterministic integration of quantum emitters and optical cavities in a van der waals crystal.
\newblock {\em arXiv preprint arXiv:2601.03803}, 2026.


\bibitem{froch2021coupling}
Johannes~E Froch, Lesley~P Spencer, Mehran Kianinia, Daniel~D Totonjian, Minh Nguyen, Andreas Gottscholl, Vladimir Dyakonov, Milos Toth, Sejeong Kim, and Igor Aharonovich.
\newblock Coupling spin defects in hexagonal boron nitride to monolithic bullseye cavities.
\newblock {\em Nano Letters}, 21(15):6549--6555, 2021.

\bibitem{flatten2018microcavity}
Lucas~C Flatten, Li~Weng, Artur Branny, Shawn Johnson, Philip~R Dolan, Aureli{\'e}n~AP Trichet, Brian~D Gerardot, and Jason~M Smith.
\newblock Microcavity enhanced single photon emission from two-dimensional wse2.
\newblock {\em Applied Physics Letters}, 112(19), 2018.

\bibitem{drawer2025tunable}
Jens-Christian Drawer, Salvatore Cianci, Vita Solovyeva, Alexander Steinhoff, Christopher Gies, Falk Eilenberger, Kenji Watanabe, Takashi Taniguchi, Ivan Solovev, Giorgio Pettinari, et~al.
\newblock Tunable ws $ \_2 $ micro-dome open cavity single photon source.
\newblock {\em arXiv preprint arXiv:2511.21630}, 2025.

\bibitem{duong2018enhanced}
Ngoc My~Hanh Duong, Zai-Quan Xu, Mehran Kianinia, Rongbin Su, Zhuojun Liu, Sejeong Kim, Carlo Bradac, Toan~Trong Tran, Yi~Wan, Lain-Jong Li, et~al.
\newblock Enhanced emission from wse2 monolayers coupled to circular bragg gratings.
\newblock {\em ACS Photonics}, 5(10):3950--3955, 2018.

\bibitem{hekmati2023bullseye}
Reza Hekmati, John~P Hadden, Annie Mathew, Samuel~G Bishop, Stephen~A Lynch, and Anthony~J Bennett.
\newblock Bullseye dielectric cavities for photon collection from a surface-mounted quantum-light-emitter.
\newblock {\em Scientific Reports}, 13(1):5316, 2023.

\bibitem{spencer2023monolithic}
Lesley Spencer, Jake Horder, Sejeong Kim, Milos Toth, and Igor Aharonovich.
\newblock Monolithic integration of single quantum emitters in hbn bullseye cavities.
\newblock {\em ACS Photonics}, 10(12):4417--4424, 2023.

\bibitem{vogl2019compact}
Tobias Vogl, Ruvi Lecamwasam, Ben~C Buchler, Yuerui Lu, and Ping~Koy Lam.
\newblock Compact cavity-enhanced single-photon generation with hexagonal boron nitride.
\newblock {\em Acs Photonics}, 6(8):1955--1962, 2019.

\bibitem{haussler2021tunable}
Stefan H{\"a}u{\ss}ler, Gregor Bayer, Richard Waltrich, Noah Mendelson, Chi Li, David Hunger, Igor Aharonovich, and Alexander Kubanek.
\newblock Tunable fiber-cavity enhanced photon emission from defect centers in hbn.
\newblock {\em Advanced Optical Materials}, 9(17):2002218, 2021.

\bibitem{maier2025extracting}
Patrick Maier and Alexander Kubanek.
\newblock Extracting membrane-like hexagonal boron nitride hosting single defect centers for resonator integration.
\newblock {\em arXiv preprint arXiv:2508.13985}, 2025.

\bibitem{proscia2020microcavity}
Nicholas~V Proscia, Harishankar Jayakumar, Xiaochen Ge, Gabriel Lopez-Morales, Zav Shotan, Weidong Zhou, Carlos~A Meriles, and Vinod~M Menon.
\newblock Microcavity-coupled emitters in hexagonal boron nitride.
\newblock {\em Nanophotonics}, 9(9):2937--2944, 2020.

\end{thebibliography}

\end{document}